\documentclass[12pt]{article}

\usepackage{amsmath,amssymb,amsfonts,graphics,graphicx,amscd,amsfonts,epsf,epsfig,color, dsfont}
\usepackage[hidelinks,pdftex]{hyperref}
\usepackage{subcaption}
\allowdisplaybreaks[4]

\date{}
\begin{document}

\begin{flushright}

\end{flushright}

\vspace{0.1cm}

\begin{center}

   {\LARGE

Partial-Symmetry-Breaking Phase Transitions
  }   
   
\end{center}
\vspace{0.1cm}
\vspace{0.1cm}

\begin{center}
Masanori Hanada$^a$
and 
Brandon Robinson$^{b,a}$
\vspace{0.5cm}

$^a${\it School of Physics and Astronomy, and STAG Research Centre}\\
{\it University of Southampton, Southampton, SO17 1BJ, UK}\\
\vspace{0.2cm}  

$^b${\it Instituut voor Theoretische Fysica, KU Leuven}\\
{\it Celestijnenlaan 200D, B-3001 Leuven, Belgium}
\vspace{0.2cm}  

\end{center}

\vspace{1.5cm}

\begin{center}
  {\bf Abstract}
\end{center}

We demonstrate a novel feature of certain phase transitions in theories with large rank symmetry group that exhibit specific types of non-local interactions.
A typical example of such a theory is a large-$N$ gauge theory where by `non-local interaction' we mean the all-to-all coupling of color degrees of freedom. 
Recently it has been pointed out that nontrivial features of the confinement/deconfinement transition are understood 
as consequences of the coexistence of the confined and deconfined phases on the group manifold describing the color degrees of freedom. 
While these novel features of the confinement/deconfinement transition are analogous to the two-phase coexistence at the first order transition of more familiar local theories, various differences such as the partial breaking of the symmetry group appear due to the non-local interaction.  
In this article, we show that similar phase transitions with partially broken symmetry can exist in various examples from QFT and string theory.    
Our examples include the deconfinement and chiral transition in QCD, 
Gross-Witten-Wadia transition in two-dimensional lattice gauge theory, 
Douglas-Kazakov transition in two-dimensional gauge theory on sphere, 
and black hole/black string transition.

\newpage
\tableofcontents

\section{Introduction}\label{sec:introduction}
\hspace{0.51cm}

In this paper, we suggest a generic feature of the phase transitions in the non-locally interacting theories with large rank symmetry group. By `non-locally interacting', in the examples that we consider in the following sections, we mean that internal degrees of freedom (e.g. colors in gauge theories) are interacting through all-to-all couplings.  Further, our analysis will be focussing primarily on theories where the interactions in spacetime are local\footnote{Note that a theory can realize local and non-local interactions simultaneously on separate abstract spaces.  A standard example -- that will comprise the starting point of most of our analysis -- would be a gauge theory understood as the fibre bundle $G\hookrightarrow E\rightarrow \mathcal{M}$: local interactions on the base (spacetime) manifold $\mathcal{M}$ and non-local interactions in the fibre $G$ are encoded in, say, the Yang-Mills action $\int_{\mathcal{M}}{\rm Tr}~ F\wedge\star F$.}. While we almost exclusively study large-$N$ gauge theories in order to realize concretely the physics of the phase transitions of interest, the logic of the underlying mechanism seems like it could be applied more broadly.

To clarify the language that we will be using to discuss phase transitions further, we will refer to the manifold on which degrees of freedom are interacting non-locally (e.g. the group manifold of $SU(N)$) as {\it internal space}. 
By {\it physical space} we are referring to spatial volumes in spacetime, and on the physical space, we assume the degrees of freedom are interacting locally.

\begin{figure}[htbp]
 \begin{center}
 \includegraphics[width=\textwidth]{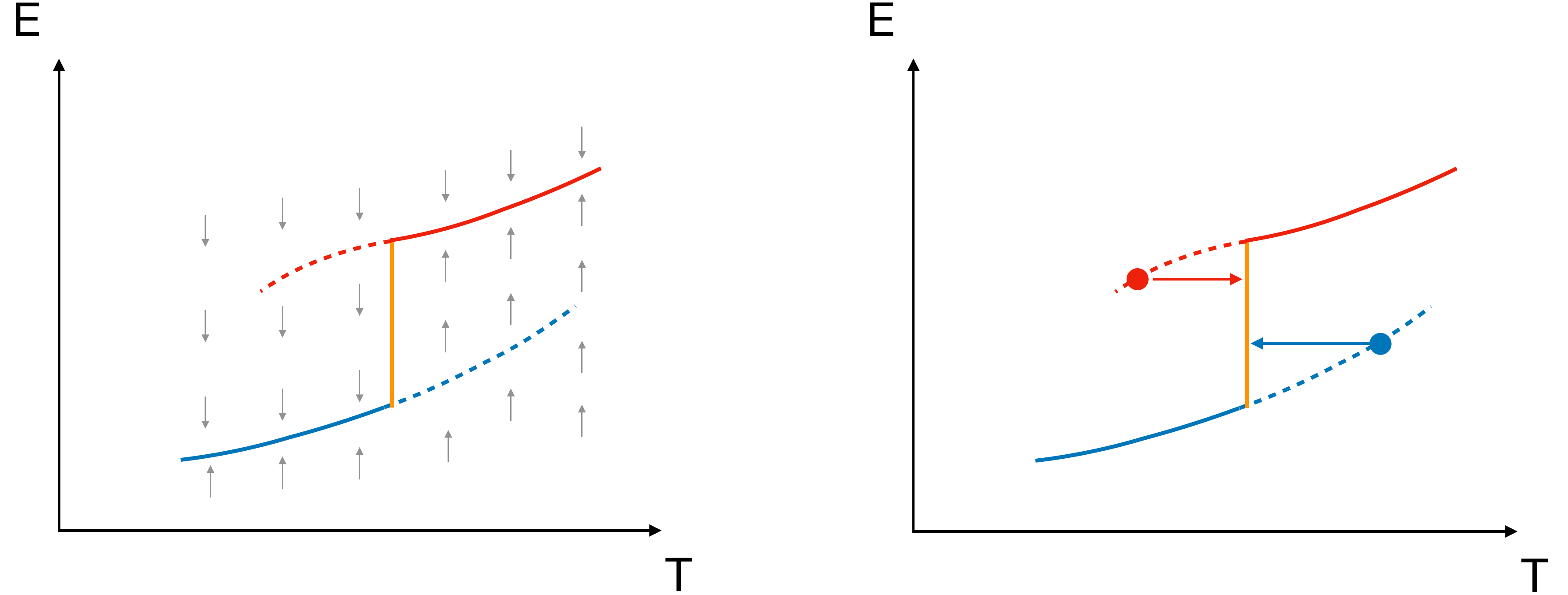}
 \end{center}
\caption{Cartoon picture of the phase diagram of liquid water and ice, in terms of the energy $E$
and temperature $T$. Red (blue) lines represents the liquid (solid) phases. 
Solid (dashed) lines indicate (meta-)stable phases. 
The vertical orange line is the liquid-solid coexistence phase. 
[Left] Canonical ensemble, i.e. $T$ is fixed and $E=E(T)$.
The solid and dashed lines represent the stable and metastable phases. 
The system moves toward the stable phase along gradients of the free energy (arrows) under small perturbations in $E$.  
[Right] Microcanonical ensemble, i.e. $E$ is fixed and $T=T(E)$.
When the metastable phases are perturbed, the system transits to the two-phase coexistent state.
These figures are taken from Ref.~\cite{Hanada:2018zxn}. 
 }\label{fig:water}
\end{figure}

As a specific example, let us consider the confinement/deconfinement transition in gauge theory \cite{Hanada:2016pwv,Berenstein:2018lrm,Hanada:2018zxn,Hanada:2019czd}, and illuminate the idea by comparing it with 
the transition between liquid and solid phases of water.  The cartoon picture of the phase diagram of water at 1-atm and near zero-celsius is shown in Fig.~\ref{fig:water}. Although, the terminology of internal and physical space in the case of analyzing the liquid-solid transition of water is a bit confusing, it will serve as a useful analog for what is happening in various phase transitions in later sections.

\begin{figure}[htbp]
\begin{center}
\begin{subfigure}[htbp]{\textwidth}
   \includegraphics[width=\textwidth]{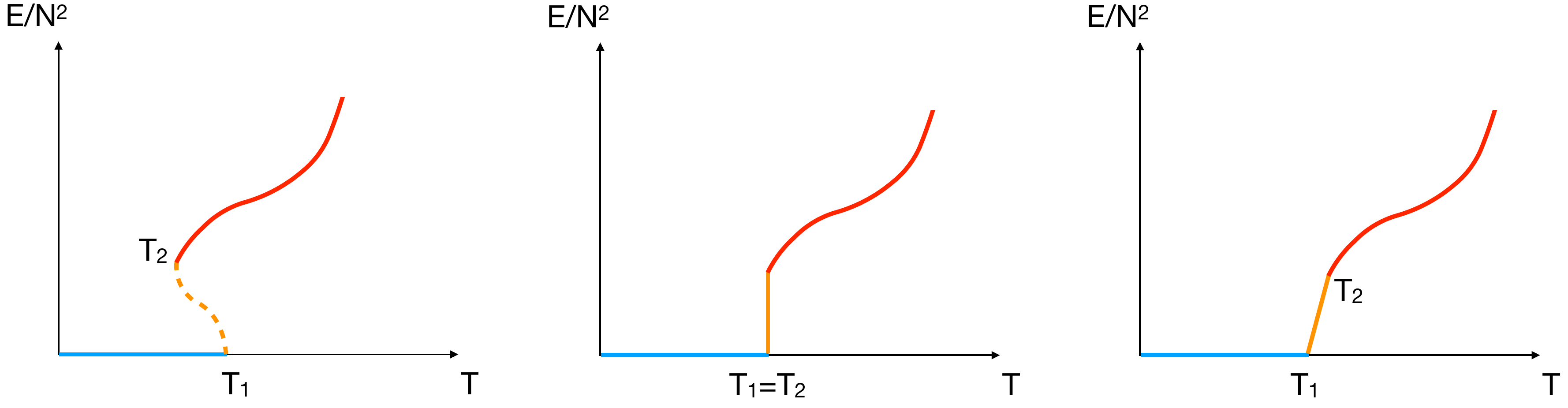}
\caption{
Three basic types of energy-vs-temperature relations in various large-$N$ gauge theories. 
 }\label{fig:E-vs-T-BH}
\end{subfigure}\newline
   \end{center}

 \begin{center}
\begin{subfigure}[htbp]{\textwidth}
  \includegraphics[width=\textwidth]{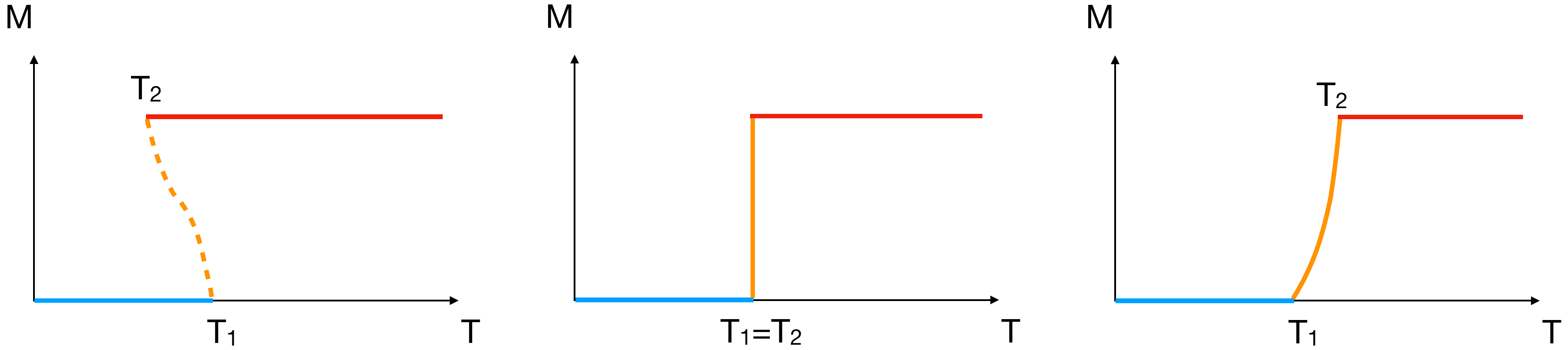}
\caption{
Three types of $M$-vs-temperature relations in various large-$N$ gauge theories. 
Here $M$ characterize the size of the deconfined phase in the color space, namely 
SU($M$) subgroup of SU($N$) is deconfined. 
 }\label{fig:M-vs-T-BH}
 \end{subfigure}
  \end{center}
  \caption{The blue (red) lines represent the confined (deconfined) phases. 
Along the orange line is the `partially deconfined' coexistence phase.
Essentially the same figures have been used in Ref.~\cite{Hanada:2018zxn}. 
}
\end{figure}

\vspace{3mm}
\noindent
{\bf Two phases can coexist in the internal space.}
At the critical temperature $T=T_{\rm c}=$ zero-celsius, water goes through a first order transition, and the liquid and solid phases can coexist
in physical space. 
The coexistence curve in the phase diagram is represented by the vertical orange line in Fig.~\ref{fig:water}. 
In a similar manner, two phases (e.g.~confined and deconfined phases) can coexist in the internal space;
see Fig.~\ref{fig:E-vs-T-BH} and Fig.~\ref{fig:M-vs-T-BH}. 
The generalization of this idea to other non-local theories is straightforward. 

\vspace{3mm}
\noindent
{\bf Partial symmetry breaking can occur.}
Associated with the coexistence of two phases in the internal space, the symmetry of the system can be broken spontaneously. 
For example SU($N$) symmetry of gauge theory 
spontaneously breaks to\footnote{
Strictly speaking, the global part of gauge symmetry can break spontaneously. For details, see Ref.~\cite{Hanada:2019czd}.
} SU($M$)$\times$SU($N-M$)$\times$U($1$).  
Moreover, the symmetry changes gradually, 
namely $\frac{M}{N}$ changes continuously from 0 to 1; see Fig.~\ref{fig:M-vs-T-BH}. 
There is no counterpart to this phenomenon in the coexistence in physical space. 
\vspace{3mm}

\vspace{3mm}

\vspace{3mm}
\noindent
{\bf Temperature can change nontrivially.}
Ignoring the interactions at the interface of the two phases, the admixture of liquid and solid water in physical space determines the energy of the configuration at fixed temperature.
On the other hand, in the non-local theory, temperature can change nontrivially, 
as shown in the left and right panels of Fig.~\ref{fig:E-vs-T-BH}. 
The same can hold for other parameters, for example when we replace the energy and temperature with charge and chemical potential.

\vspace{3mm}
\noindent
{\bf `Metastable' phases can be completely stabilized in the thermodynamic limit.}
Associated with the first order transition, metastable phases such as supercooled liquid and superheated solid 
can appear. Such metastable phases due to the local interaction are unstable to a small perturbation, even in the thermodynamic (large volume) limit. 
On the other hand, similar `metastable' phases due to the non-local interaction can be stabilized in the thermodynamic limit (the large-$N$ limit), namely the size of the perturbation needed to destabilize such phases can increase with the system size. 
While this fact has been well known for the large-$N$ gauge theories, the apparent difference from the local interaction has not been widely appreciated. 
\vspace{3mm}

In the rest of the paper, we collect evidence for these statements, 
based on concrete examples and heuristic arguments which apply to generic cases. 
We will start with qualitative arguments in Sec.~\ref{sec:qualitative_argument},\footnote{
This section rephrases the statements in Refs.~\cite{Hanada:2018zxn,Hanada:2019czd} in a more intuitive manner,  
so that the generalization to other theories becomes straightforward. 
} where we discuss two seemingly drastically different examples --- the brane model and the ant model --- 
which share many common features. 
Following on that discussion, we will outline a heuristic proposal for extending the lessons learned from the brane and ant models to more generic cases.
In Sec.~\ref{sec:QCD}, we will study QCD-like theories on a compact physical space (i.e. S$^3$) in various limits giving both analytic and numerical quantitative evidence for the appearance of partial symmetry breaking phase transitions, and at the end we will explore the possible effects that partial deconfiment could have on chiral symmetry.  In Sec.~\ref{sec:2d-YM}, we will analyze partial-symmetry-breaking phase transitions in 2d pure Yang-Mills (YM) theories in both lattice and continuum formulations.  In Sec.~\ref{sec:BH-BS-2d-YM} we will try to understand partial symmetry breaking phase transitions through gauge/gravity duality in the context of black hole/black string transitions and their relation to maximally supersymmetric 2d super-Yang-Mills (SYM) theory.
In Sec.~\ref{sec:R3*S1_pbc}, the Higgsing in QCD with adjoint quarks is discussed. 
Finally, we will provide a brief recap, discussion, and outlook for future work.

\section{Qualitative/Phenomenological examples}\label{sec:qualitative_argument}
\hspace{0.51cm}

Before moving on to more complicated, quantitative descriptions, let us start exploring the ideas discussed above with a set of simple phenomenological models that capture broad features of the physics at large $N$ phase transitions that we wish to argue are universal.
The readers who are not familiar with superstring theory can skip Sec.~\ref{sec:D-brane}.

\subsection{Brane model}\label{sec:D-brane}
\hspace{0.51cm}
Let us consider the system of $N$ coincident D-branes \cite{Dai:1989ua}
and open strings connecting them: The worldvolume theory of this collection of D-branes has a low-energy description as supersymmetric Yang-Mills (SYM) theory with an SU($N$) gauge symmetry \cite{Witten:1995im}. 
As there is a gauge theoretic description of the dynamics of the worldvolume theory, the interactions on the internal space (of color degrees of freedom in the low-energy SYM theory description) are non-local, while the physical space is local. 

Suppose we separate out $M(<N)$ D-branes and form a bound state, which is interpreted as black hole; then in total, $\mathcal{O}(M^2$) open string degrees of freedom are excited, and hence, the extensive variables (e.g. entropy and energy) are of order $\mathcal{O}(M^2)$. 
If another D-brane approaches the bound state, $M$ different open string modes can be excited and pull the additional brane. The attractive force between the bound state and additional brane is easily interpreted as entropic force: if the size of the bound state increases to $M+1$, entropy increases from $\sim M^2$ to $\sim (M+1)^2$. Thus, there is an entropic force $\sim M$. 
At the same time, because the energy increases from $E\sim M^2$ to $E\sim (M+1)^2$,
the suppression due to the Boltzmann factor $e^{-E/T}$ tends to push down the value of $M$. 
The size of the bound state is determined so that these two contributions balance. 
In this way, {\bf two phases can coexist in the internal space} of the worldvolume theory; namely $N$ D-branes splits to two groups, $M$ inside and $N-M$ outside the bound state. 

In the large $N$ limit, as the value of $M$ increases from $0$ to $N$, $\frac{M}{N}$ changes continuously. In the bound state phase, the worldvolume gauge symmetry is broken to SU($M$)$\times$SU($N-M$)$\times$U(1).\footnote{
If the flat direction is allowed, SU($M$)$\times$U(1)$^{N-M-1}$ can also be realized.} 
In this way, {\bf partial symmetry breaking can occur}.  
There are two phase transitions: $\frac{M}{N}=0$, where the bound state formation begins, 
and $\frac{M}{N}=1$, where all of the branes have fallen behind the horizon of the black hole formed by the bound state.

Let us make the argument more precise. Consider the canonical ensemble at finite temperature $T$. 
The partition function can be written as 
\begin{eqnarray}
Z(T)
=
\int dE \Omega(E)e^{-E/T}
=
\int dE e^{-F(E,T)/T}. 
\end{eqnarray}
Here we denote the density of states at energy $E$ by $\Omega(E)=e^{S(E)}$, where $S(E)$ is the entropy.
The Helmholtz free energy $F(E,T)$ is given by the standard expression
\begin{eqnarray}
F(E,T)
=
E-TS(E). 
\end{eqnarray}
Extremizing the free energy with respect to the energy and utilizing basic fact in the microcanonical ensemble $\frac{dS}{dE} = T_{\rm micro}^{-1}$, where $T_{\rm micro}$ is the microcanonical temperature, one can easily see that $\frac{\partial F}{\partial E}=0$ is equivalent to setting $T=T_{\rm micro}$. 
Therefore, by finding the saddle points for fixed canonical temperature $T$, we obtain the energy $E$ which corresponds 
to the microcanonical temperature $T_{\rm micro}=T$.

After finding the saddle points, we can characterize the stability of the solutions to $\frac{\partial F}{\partial E}=0$ by computing $\frac{\partial^2 F}{\partial E^2}>0$  (stable) or 
$\frac{\partial^2 F}{\partial E^2}<0$ (unstable). 
Equivalently, by using
\begin{eqnarray}
\frac{\partial^2 F}{\partial E^2}
=
\frac{T}{T_{\rm micro}^2}\cdot\frac{d T_{\rm micro}}{dE},  
\end{eqnarray}
we can identify stable and unstable saddles corresponding to the phases with positive specific heat and negative specific heat, respectively. 

In Fig.~\ref{fig:E-vs-T-BH} and Fig.~\ref{fig:M-vs-T-BH}, we have shown the cartoon pictures of three basic patterns of
the phase diagram \cite{Sundborg:1999ue,Aharony:2003sx, Hanada:2018zxn}.  
The middle panel corresponds to the weak-coupling limit of pure Yang-Mills theory \cite{Sundborg:1999ue,Aharony:2003sx} with 
the intermediate phase represented by the vertical orange line. In this case, the saddle corresponding to the intermediate phase has $\frac{\partial^2 F}{\partial E^2} = 0$ regardless of the value of $E$.
In the left and right panels, the specific heat of the intermediate phase is negative and positive, respectively. 
The reason that nontrivial temperature dependence appears can be understood as follows \cite{Berkowitz:2016znt,Hanada:2016pwv}:
Roughly speaking, temperature is energy per number of dynamical degrees of freedom,
and hence, $T\sim E/M^2$.\footnote{In the brane model, the energy is dominated by the contribution from the bound state.} 
Note that $M^2$ is parametrically determined as a function of $E$, and hence, the specific heat can be positive or negative depending on the detail of the dynamics. In this way, {\bf temperature can change nontrivially.}

When the phase diagram in the canonical ensemble is like the left panels of Fig.~\ref{fig:E-vs-T-BH} and Fig.~\ref{fig:M-vs-T-BH},
the transition becomes of first order with hysteresis.
Typically, free energies of two stable saddles are not equal, and hence, one of them is just metastable. 
However, in order for the tunneling to happen, it is necessary to go beyond the unstable saddle,
and hence a very large perturbation is needed.  That is, the value of the size of the bound state $M$ and free energy $F$ 
have to change the amount of $\mathcal{O}(N)$ and $\mathcal{O}(N^2)$, respectively, 
and the tunneling rate becomes $\sim e^{-N^2}$.  
In this way,  {\bf `metastable' phase can be completely stabilized in the thermodynamic limit (large-$N$ limit).}

\subsection{Ant trail formation}
\hspace{0.51cm}
Departing entirely from high energy physics, let us consider a somewhat funny example from entomology: the formation of an ant trail \cite{Beekman9703}. 
This cute, relatively simple, phenomenological model has a phase behavior that strikingly resembles the brane model \cite{Hanada:2018zxn}.  

To summarize the ant model in \cite{Beekman9703}, consider a colony of $N$ indistinguishable ants such that there exists an S$_N$ permutation symmetry exchanging any individual member of the colony.  The number of ants forming a trail as a function of time is denoted $M(t)$ and depends on a number of environmental factors -- e.g. local topography, weather, and so on -- such that the time dependence of $M$ is modeled as
\begin{eqnarray}
\frac{dM}{dt}
&=&
(\alpha+pM)(N-M)
-
\frac{sM}{s+M}.
\label{eq:ant-equation}
\end{eqnarray}
The parameters in eq.~\eqref{eq:ant-equation} encode the system dynamics as: $\alpha$ speciying the probability that each ant find the food source accidentally, $s$ describing the rate that ants leave the trail, and $p$ gives the strength of the pheromone-mediated non-local two-body interaction between each ant. To give a sense of the environmental dependence in the model, consider a day with particularly dry air conditions: the lack of ambient moisture in the air the pheromone evaporates faster and hence $p$ becomes smaller. 
The size of the stationary ant trail can be obtained by solving the equation $\frac{dM}{dt}=0$. 

While it is true that $M$ can change continuously, some care is needed with regards to the distinction 
between canonical and microcanonical ensemble.  We follow the treatment in Ref.~\cite{Hanada:2018zxn}, 
in order to make the application to physics problems more salient.  
That is, there are analogous features shared between the stationary trail equation and the saddle point equation for other systems we discuss later.

\begin{figure}[htbp]
 \begin{minipage}{0.32\hsize}
 \begin{center}
   \scalebox{1.3}{\hspace{-0.45cm}
  \includegraphics[width=50mm]{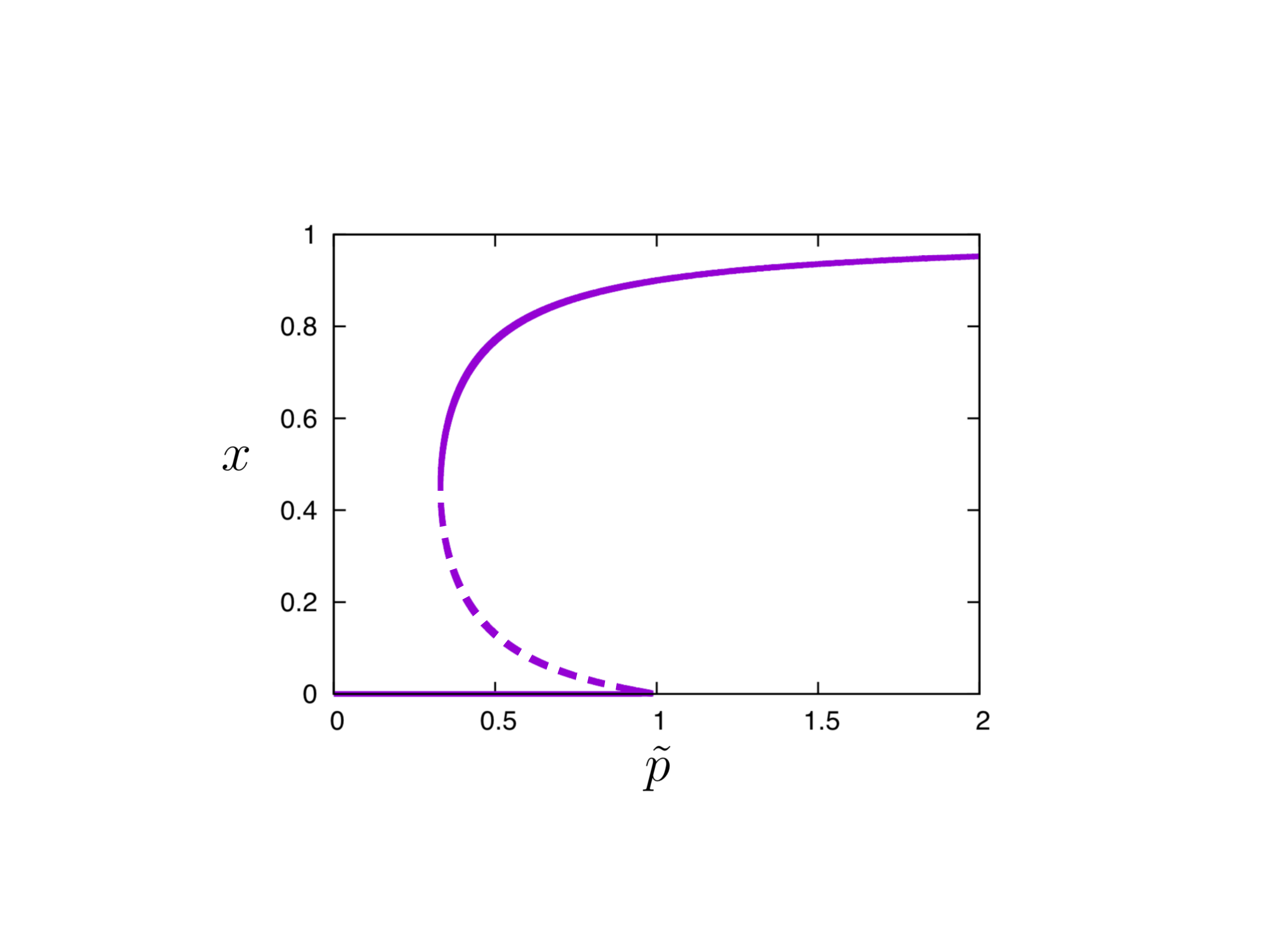}}
 \end{center}
 \end{minipage}
 \begin{minipage}{0.32\hsize}
 \begin{center}
   \scalebox{1.3}{\hspace{-0.45cm}
  \includegraphics[width=50mm]{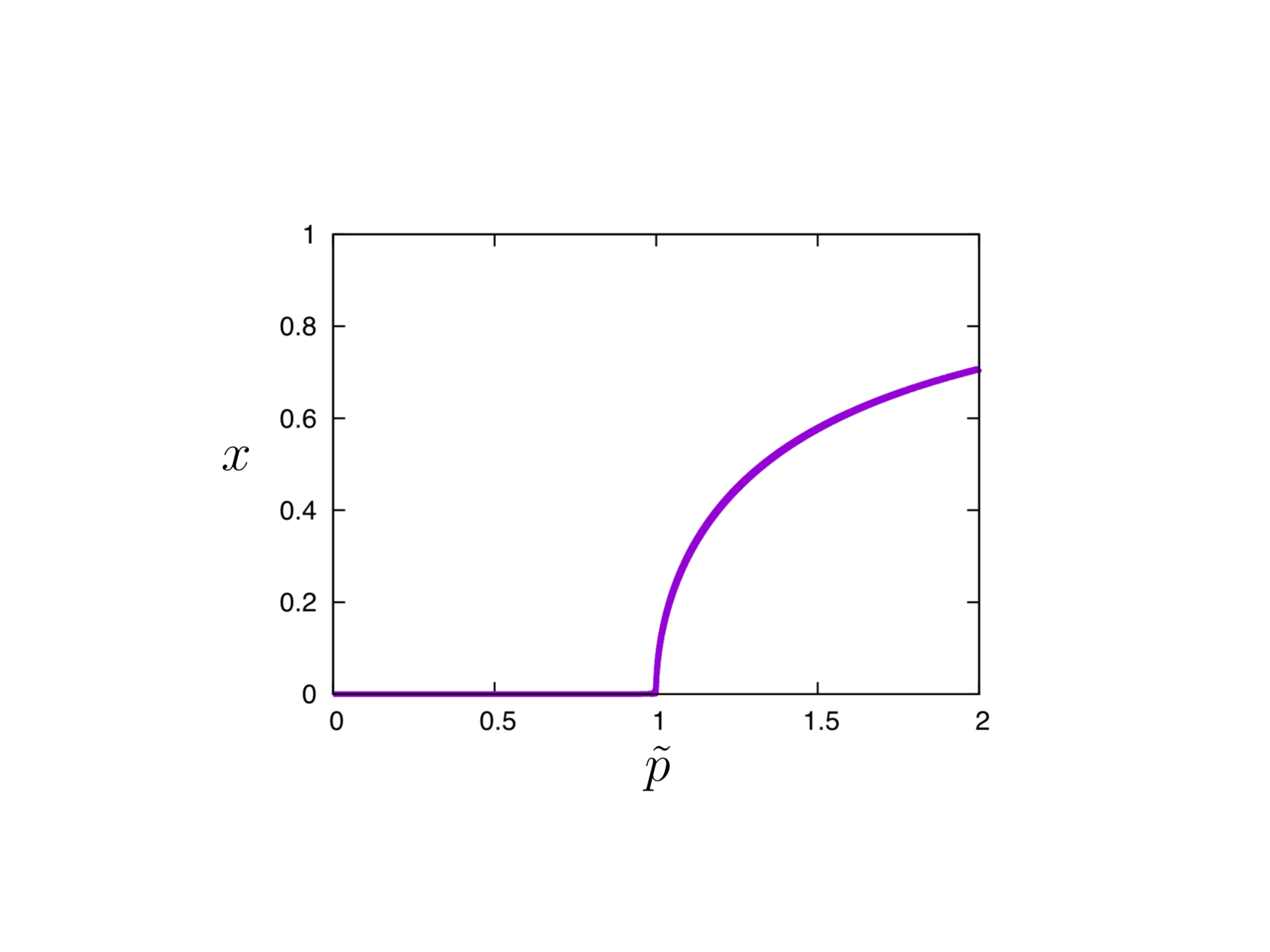}}
 \end{center}
 \end{minipage}
  \begin{minipage}{0.32\hsize}
 \begin{center}
   \scalebox{1.3}{\hspace{-0.45cm}
  \includegraphics[width=50mm]{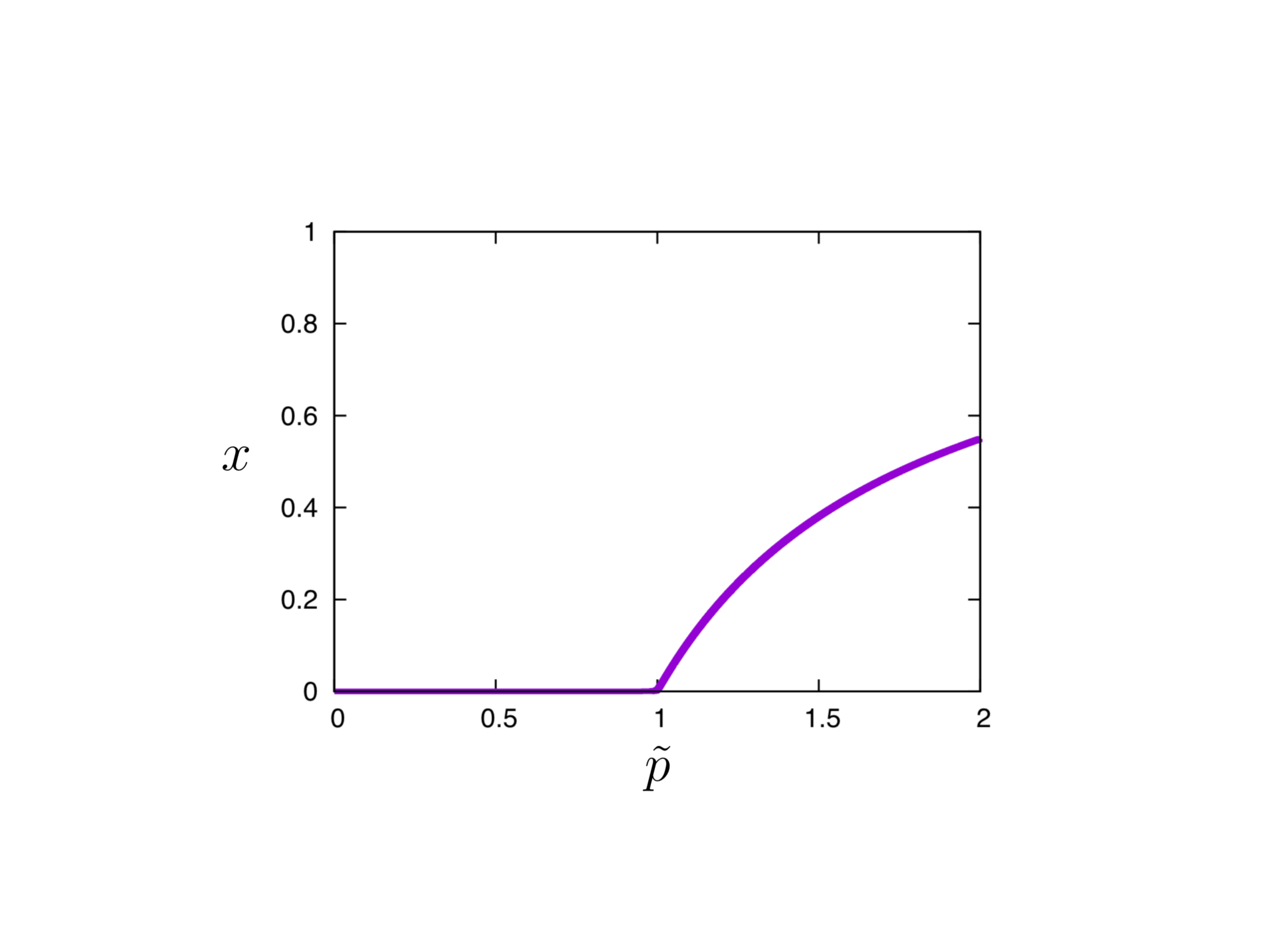}}
 \end{center}
 \end{minipage}
\caption{
$x\equiv\frac{M}{N}$ versus $\tilde{p}\equiv Np$ in the ant trail model \eqref{eq:ant-equation}. 
$N\alpha=1$, $\tilde{s}\equiv\frac{s}N=0.1$ (left), $1.0$ (center) and $5.0$ (right), 
$N=10^5$. 
Pictures taken from Ref.~\cite{Hanada:2018zxn}.  
 }\label{fig:Ntrail-vs-p}
\end{figure}

An interesting `large-$N$' limit of the ant model is taken where
$\alpha\sim N^{-1}$, $p\sim N^{-1}$, $s\sim N^1$ \cite{Hanada:2018zxn}. 
In Fig.~\ref{fig:Ntrail-vs-p}, we show the plot of $x\equiv\frac{M}{N}$ versus $\tilde{p}\equiv Np$ 
for $N\alpha=1$, several $\mathcal{O}(1)$ values of $\tilde{s}\equiv\frac{s}N$, and $N=10^5$.
As we will explain later, $\tilde{p}$ is analogous to temperature in the brane model \cite{Hanada:2018zxn}. 
The number of the ants $M$ corresponds to the number of D-branes in the bound state. 

In the three examples of varying $\tilde{s}$ increasing from left to right in Fig.~\ref{fig:Ntrail-vs-p}, we see the stable -- or possibly metastable -- saddles (solid lines) and unstable saddles (dashed lines) displaying interesting phase behavior.
Whether the saddle is stable or unstable can be seen from the sign of $\frac{dM}{dt}$ around the saddle. 
Note that here we vary $\tilde{p}$ and study the response in $M$, which amounts to being in the canonical ensemble. Starting from the leftmost panel in Fig.~\ref{fig:Ntrail-vs-p}: 

\begin{itemize}
\item
For $\tilde{s}<1$ in the left panel of Fig.~\ref{fig:Ntrail-vs-p}, if under small perturbations to $M$ away from the saddle the trail accumulates new ants or disintegrates, the saddle is unstable (dashed line). This unstable saddle separates two (meta-)stable saddles and causes the first order phase transition. 

In order for the tunneling between two stable saddles to happen, it is necessary to go beyond the unstable saddle.
It requires $\mathcal{O}(N)$ change of the number of the ants, and hence, 
as $N$ grows it becomes harder and harder to go across this unstable phase.
Therefore, there is no tunneling in the large-$N$ limit.   

\item
When $\tilde{s}>1$ in the right panel of Fig.~\ref{fig:Ntrail-vs-p}, under small perturbations to $M$, the saddle is stable at all $\tilde{p}$. 

\item
At $\tilde{s}=1$ in the center panel of Fig.~\ref{fig:Ntrail-vs-p}, we see the boundary between two cases where
the curve $x = x(\tilde{p})$ has infinite slope at $\tilde{p}=1$ at large $N$.

\end{itemize}

It is easy to find the analogy to the system of D-branes \cite{Hanada:2018zxn}. 
At higher temperature, each open string mode can be excited more, and hence the attraction becomes stronger: 
The analog in ant trail formation is that the non-local coupling (pheromone) attracting ants to the trail becomes stronger. 
The suppression of $M$ due to the Boltzmann factor balancing the entropic force is played by the `laziness' term 
in the ant model, which describes the rate of the ants leaving the trail.
Taking the analogy to its end point, just as the balance in entropic (open string) forces and Boltzmann suppression allowed for the formation of a D-brane bound state, a stationary ant trail phase is formed when the inflow and outflow balance.

The internal space consisting of the ants splits into two simultaneously realized phases: inside and outside the trail. 
Hence two phases can coexist in the internal space. 
The permutation symmetry of the ants breaks as $S_N\to S_M\times S_{N-M}$, and hence  
partial symmetry breaking can occur. 
The analogue of temperature is the pheromone parameter $\tilde{p}$.
As we can see from Fig.~\ref{fig:Ntrail-vs-p}, 
$\tilde{p}$ can change nontrivially as a function of $x$. 
Finally, in the left panel of Fig.~\ref{fig:Ntrail-vs-p}, 
$\mathcal{O}(N)$ perturbation to the number of ants on the trail is needed in order for the `tunneling' across the unstable trail to happen, 
and hence, 
`metastable' phase can be completely stabilized in the thermodynamic limit (many-ant limit). 

\subsection{Underlying mechanism}
\hspace{0.51cm}
Having constructed two rather simple models exhibiting partial-symmetry-breaking phase transitions, let us understand the mechanism behind their phase diagrams 
and extract the lessons applicable to other systems as well. 

Returning to the liquid/solid transition of water explained in Sec.~\ref{sec:introduction}, the two phases have different energy due to finite latent heat. In the microcanonical ensemble, there is no discontinuity because the energy of the entire system and the volume occupied by each phase can continuously change in a correlated manner.  As we will see in the following, essentially the same mechanism works in the internal space of non-locally interacting systems. 

Typically in the large-$N$ limit, $\mathcal{O}(N)$ degrees of freedom interact non-locally on the internal space defined by the symmetries of the theory. Further in the large-$N$ limit, phase transitions exhibit a large jump in the extensive thermodynamic variables in the canonical ensemble, e.g. the energy or charge change significantly.
In the microcanonical description, however, one expects that there is path in phase space describing a (possibly unstable) intermediate phase smoothly connecting the endpoints of the jump. The partially broken symmetries in the theory at points on the path through the intermediate phase then smoothly interpolate between the symmetries in the (meta-)stable phases, e.g. the S$_M\times$S$_{N-M}$-phase in the ant-model or SU($M$)$\times$SU($N-M$)$\times$U(1) phase for the D-brane picture of black hole formation.  
 
This raises a question in the intermediate phase: What if, e.g., all $N$ D-branes formed a bound state or all $N$ ants formed a trail immediately at the phase transition, and all degrees of freedom on the internal space were excited simultaneously at small energy? 
Then temperature, which is roughly $\sim E/N^2$, would have to become parametrically small, 
and hence the microcanonical phase diagram would have a discontinuity.  
Therefore, as the function of $E$, the number of unlocked degrees of freedom $M^2$ has to increase gradually, so that $E/M^2$ is continuous.\footnote{
One may think the continuity can be achieved if small fraction in the spatial volume goes thorough complete deconfinement. 
However, note that the deconfinement takes place even in matrix models, in which there is no spatial dimension at all. 
See Sec.~\ref{sec:large-volume-vs-small-volume} for more explanation regarding this issue. 
} A similar way to phrase it is: in quantum mechanics, it is impossible to excite all degrees of freedom in parametrically small amount on a compact space because each mode is quantized. 

The continuous change of the symmetry --- here we consider an SU($N$) symmetry group ---  in generic cases cannot be concluded by the continuity of the temperature alone; \textit{a priori}, $M^2$ degrees of freedom may or may not be organized in representations of, e.g., an SU($M$) subgroup.  
In order to make clear the reason that we conjecture the continuous change of the symmetry, 
let us rephrase the statement as follows: although full SU($N$) symmetry cannot be preserved due to the $M^2$ excitations, the system still tends to keep as large symmetry as possible. 
We conjecture this because it is natural to expect that the solution to the extremization problems (the maximization of the entropy in the case of the microcanonical ensemble) respects the symmetry as much as possible. 

Essentially the same can happen in other systems as well. The point is that there are many degrees of freedom interacting non-locally, 
and they can be separated to two phases in the `internal space'. 
The continuity in the microcanonical ensemble requires the gradual change of the numbers of degrees of freedom in two phases. 
The same principle --- the continuity in the microcanonical ensemble --- naturally leads to the intermediate phase, or equivalently the coexistence of two phases, both in non-locally and locally interacting theories.
Furthermore, when the system has a large symmetry, it is natural to expect the continuous change of the symmetry.  

An important difference is how $M$ in the non-local theories --  
or the volume of each phase in the local theories -- depends on the energy $E$. 
In the local theories, as long as the contribution from the interface is negligible, 
the volume depends linearly on the energy, analogous to the center panel of Fig.~\ref{fig:M-vs-T-BH}. 
In the non-local theories, more complicated behavior can show up. 

When the specific heat of the intermediate phase is negative,
it becomes the maximum of free energy in canonical ensemble, 
which separates stable and metastable phases.  
The amount of fluctuation needed for the tunneling to the other (meta-)stable branches increases with $N$. 
Therefore, in the thermodynamic limit (large-$N$ limit), the probability of seeing a tunneling event becomes vanishingly small
and the `metastable' state is completely stabilized at $N=\infty$. 

All of the ingredients explained above are not specific to the deconfinement transition in gauge theory, and so it is a reasonable conjecture that there is a large universality class of theories that exhibit such phase behavior.

\subsubsection{Large volume vs small volume}\label{sec:large-volume-vs-small-volume}
\hspace{0.51cm}
Often, a theory is described in terms of a volume in physical space and a phase on internal space simultaneously. 
If the physical volume is so small that the coexistence of two phases in physical space is impossible, 
we need to consider only the coexistence in the internal space. 
What if both internal and physical spaces are large?  

Let us consider microcanonical ensemble, and
suppose we prepare an initial state in the intermediate phase with uniform energy density. 
Locally, there are three basic patterns as we have already seen (Fig.~\ref{fig:E-vs-T-BH} and Fig.~\ref{fig:M-vs-T-BH}), 
as in the case of the small volume.  
If the specific heat in the intermediate phase is negative, fluctuations in the internal space degrees of freedom can lead to instability in the physical space.\footnote{
Consider two small neighboring volumes $A$ and $B$ in thermodynamic equilibrium, and a small amount of energy moved from $A$ to $B$. 
After the energy transfer, the temperature in $A$ will be slightly higher than that in $B$ because the specific heat is negative. 
Following the energy transfer, the resulting gradient of the temperature induces further energy transfer, and thus, after a small perturbation the difference of the temperatures in regions $A$ and $B$ grows. 
} 
Typically, a spinodal decomposition takes place and most of the physical space is occupied 
by the phases with either $\frac{M}{N}=0$ or $\frac{M}{N}=1$. Yet, the intermediate phase $0<\frac{M}{N}<1$ survives on the domain wall separating regions existing in the two (meta-)stable phases.
On the other hand, if the specific heat coming from the dynamics in the internal space is positive, 
then such instability does not arise and the intermediate state can stably fill whole the physical space. 

Let us consider further on the case with the negative specific heat (the left panels of Fig.~\ref{fig:E-vs-T-BH} and Fig.~\ref{fig:M-vs-T-BH}). 
Treating the system now in the canonical ensemble, the phase with either $\frac{M}{N}=0$ or $\frac{M}{N}=1$ filling whole the physical space gives the minimum of free energy. However, depending on the temperature, these saddles might be only local minima. 
Whether tunneling to the global minimum takes place or not depends on the order of limits in taking both large-$N$ and large physical volume: That is, fixing $N$ large but finite and taking a large volume limit may see different tunneling rates as compared to the limit taking large but fixed volume and $N\rightarrow \infty$. 
For example for a given region at fixed volume in physical space, taking the large-$N$ limit parametrically suppresses the probability of any tunneling to occur, e.g. tunneling rates in the gauge theories are typically $\mathcal{O}(e^{-N^2})$ suppressed: However, in fixing $N$ large but finite, the large physical volume limit will eventually surpass the large-$N$ suppression and allow for a non-negligible tunneling probability.

\section{QCD phase transition}\label{sec:QCD}
\hspace{0.51cm}
In this section, we will consider the coexistence of confined and deconfined phases in large $N_c$ QCD with SU($N_c$) gauge group and $N_f$ quarks in two different limits.  
Firstly, we will see that, in the analytically solvable regime of weakly coupled large $N_c$ QCD on a small S$^3$, an explicit construction of the partial deconfinement transition is possible.  
Secondly, we will study the consequences of partial deconfinement for chiral symmetry in the infinite volume limit of large $N_c$ QCD with $N_f$ massless quarks.

\subsection{Weakly-coupled QCD on S$^3$}\label{sec:free-QCD}
\hspace{0.51cm}

Let us consider SU($N_c$) YM with $N_f$ fermions of mass $m$ in the fundamental representation of the gauge algebra $\mathfrak{su}(N_c)$, and take large-$N_c$ with $\frac{N_f}{N_c}$ fixed. We will put this theory on S$^3 \times$ S$^1$ where the S$^3$ radius is $R$ and S$^1$ radius is $\beta$, the weak-coupling limit can be solved analytically in the regime where $\frac{1}{R}\gg\Lambda_{\rm QCD}$ \cite{Sundborg:1999ue,Aharony:2003sx}\cite{Schnitzer:2004qt,Hollowood:2012nr}.

\subsubsection{$N_f=0$}
\hspace{0.51cm}
Let us start with a simpler case with $N_f=0$ \cite{Hanada:2019czd}. 
To study the analytically solvable regime further, let us fix the gauge so that the temporal component of gauge field $A_t$ is static and diagonal.
In this gauge, all of the spatial components of the gauge field can be integrated out leaving behind an effective action for $A_t$ in terms of $u_n=\frac{1}{N}{\rm Tr}{\cal P}^n$.   Here we have denoted the Polyakov line made of $A_t$ as ${\cal P}$.  The $u_n$'s are determined such that the effective action is minimized.
The minimum of the effective action is interpreted as $\beta F$, where $F$ is the free energy of the theory and  $\beta = 1/T$ is the inverse temperature.
We use the same notation $\beta F$ both for the effective action and free energy, with the understanding that 
the former is the function of $u_n$'s and the latter is obtained by taking the saddle point value. 
Explicitly computing the effective action, one obtains the effective action 
\begin{eqnarray}
\beta F
=
N_c^2
\sum_n
a_n(T)
u_n^2, 
\label{YM-S3-free-energy}
\end{eqnarray}
where 
\begin{eqnarray}
a_n(T)
=
\frac{1}{n}
\left(
1
-
2\sum_{l=1}^\infty
l(l+2)e^{-n\frac{\beta (l+1)}{R}}
\right).  
\label{eq:a_n}
\end{eqnarray}

In the following, we will be concerned primarily with the thermodynamic profile of loop operators built out of $\mathcal{P}$.  
The phase distribution of the Polyakov line is given by 
\begin{eqnarray}
\rho(\theta)=
\frac{1}{2\pi}
\left(
1+2\sum_nu_n\cos(n\theta)
\right), 
\label{SUNc-Polyakov-phase}
\end{eqnarray}
where $-\pi\le\theta<\pi$. 
At low temperature, all $a_n$'s are positive, and hence all $u_n$'s take zero. 
This is the confining phase, $\rho(\theta)=\frac{1}{2\pi}$. 
The phase transition happens at $T=T_c$, where $a_1$ becomes zero:
\begin{eqnarray}
a_1(T_c)=0. 
\end{eqnarray}
At this point, other $a_n$'s ($n\ge 2$) are still positive. 
Therefore, $u_1$ can take any value without changing free energy, while other $u_n$'s remain zero, as long as $\rho(\theta)\ge 0$ is not violated. 
The distribution of the Polyakov line phase becomes
\begin{eqnarray}
\rho(\theta)
=
\frac{1+2u_1\cos\theta}{2\pi}
=
\frac{1+2P\cos\theta}{2\pi},  
\end{eqnarray}
where $P=\frac{1}{N}{\rm Tr}\cal{P}$, 
and hence, $u_1=P$ can take any value between $0$ and $\frac{1}{2}$.  
Above $T_c$, other $u_n$'s take nonzero values as well, such that the effective action is minimized without violating $\rho(\theta)\ge 0$. 

The energy at $T=T_c$ is\footnote{
By definition, $E = -\partial (\beta F)/\partial \beta$. 
In order to use this, we need to introduce a small interaction so that the $\beta$-derivative is mathematically well-defined.  
For example, we can use the expression for non-zero $\frac{N_f}{N_c}$ \eqref{QCD-S3-energy-lowT} and then send $\frac{N_f}{N_c}$ to zero. 
} 
\begin{eqnarray}
E=N_c^2P^2\times \left.\frac{\partial a_n}{\partial\beta}\right|_{T=T_c}. 
\end{eqnarray}
The entropy is obtained as $S=\beta(E-F)$, which is simply $\beta E$ at $T=T_c$. 

Therefore, the phase diagram becomes like the center panels of Fig.~\ref{fig:E-vs-T-BH} and Fig.~\ref{fig:M-vs-T-BH}. 
In the microcanonical ensemble, there are two transitions, at $P=0$ and $P=\frac{1}{2}$. The latter is the Gross-Witten-Wadia (GWW) transition \cite{Gross:1980he,Wadia:2012fr}, 
which corresponds to the formation of the gap in the phase distribution. 

We identify the GWW transition with the transition from partially deconfined phase 
to completely deconfined phase. 
The energy and entropy of the SU$(M)$-deconfined phase should be those of the SU($M$) theory at the GWW point \cite{Hanada:2016pwv,Hanada:2018zxn}: 
\begin{eqnarray}
S=S_{\rm GWW}(M) , 
\label{eq:entropy}
\end{eqnarray}
\begin{eqnarray}
E=E_{\rm GWW}(M), 
\label{eq:energy}
\end{eqnarray}
These relations actually hold, with the following identification: 
\begin{eqnarray}
P=\frac{M}{2N}.  
\label{free-YM-P-vs-M}
\end{eqnarray} 

Another nontrivial relation is obtained by looking at the distribution of Polyakov loop phases. 
Because $N-M$ of the phases are in the confining phase while the other $M$ are at the GWW point of the SU($M$) theory, 
we naturally expect \cite{Hanada:2018zxn}
\begin{eqnarray}
\rho(\theta)
&=&
\left(
1
-
\frac{M}{N}
\right)
\rho_{\rm confine}(\theta)
+
\frac{M}{N}
\cdot
\rho_{\rm GWW}(\theta; M)
\nonumber\\
&=&
 \frac{1}{2\pi}
\left(
1
-
\frac{M}{N}
\right)
+
\frac{M}{N}
\cdot
\rho_{\rm GWW}(\theta; M). 
\label{eq:Polyakov_loop_partial_deconfinement}
\end{eqnarray}
Here $\rho_{\rm GWW}(\theta; M)$ is the distribution in the SU($M$) theory at the GWW point. 
This relation holds as well, with the identification \eqref{free-YM-P-vs-M}
and $\rho_{\rm GWW}(\theta; M)=\frac{1+\cos\theta}{2\pi}$.  

We can push the argument further and demonstrate partial deconfinement more robustly by constructing the states in the Hilbert space \cite{Hanada:2019czd}. 

\subsubsection{$N_f>0$}
\hspace{0.51cm}
Let us move on to the case of $N_f>0$. 
The effective action is \cite{Schnitzer:2004qt,Hollowood:2012nr}
\begin{eqnarray}
\beta F
=
\sum_n\left(
N_c^2a_n(T)
u_n^2
+
N_cN_f
b_n(T)u_n
\right),  
\label{QCD-S3-free-energy}
\end{eqnarray}
where $a_n$ is given by \eqref{eq:a_n}, and 
\begin{eqnarray}
b_n(T)
=
\frac{(-1)^n}{n}\cdot
4\sum_{l=1}^\infty
l(l+1)e^{-n\frac{\beta}{R}\sqrt{(l+\frac{1}{2})^2+m^2R^2}}. 
\end{eqnarray}

In the following, we will be concerned primarily with the thermodynamic profile of loop operators built out of $\mathcal{P}$ again.  By definition, the phase distribution of the Polyakov loop given in 
\eqref{SUNc-Polyakov-phase}
must be positive semi-definite: $\rho(\theta)\ge 0$. 
Further in the low temperature regime $\beta \gg R$, the phase distribution $\rho(\theta)$ is everywhere nonzero. 
However, there is a GWW phase transition, which we denote by $T=T_{\rm GWW}$, above which
$\rho(\theta)$ is zero for a part of the range of $\theta$. 

As long as this phase distribution is everywhere positive, $u_n$'s can change independently, and the saddle point equation $\frac{\partial (\beta F)}{\partial u_n}=0$ can be solved for each $n$ as 
\begin{eqnarray}
u_n
=
-
\frac{N_f}{N_c}\cdot\frac{b_n(T)}{2a_n(T)}. 
\label{QCD-S3-saddle-point}
\end{eqnarray}
Evaluating the effective action on the saddle, i.e. substituting \eqref{QCD-S3-saddle-point} into \eqref{QCD-S3-free-energy}, we obtain
\begin{eqnarray}
\beta F
=
-\frac{N_f^2}{4}
\sum_n
\frac{\left(b_n(T)\right)^2}{a_n(T)}. 
\label{QCD-S3-free-energy-2}
\end{eqnarray}
From this point, computing the internal energy of the thermodynamic system ($E = -\frac{\partial (\beta F)}{\partial \beta}$) yields
\begin{eqnarray}
E &=&
\frac{N_f^2}{4}
\sum_n\left\{
\left(\frac{b_n}{a_n} \right)^2
\cdot
\frac{\partial a_n}{\partial\beta}
-
2
\frac{b_n}{a_n}
\cdot
\frac{\partial b_n}{\partial\beta}
\right\}
\nonumber\\
&=&
\sum_n\left(
N_c^2u_n^2\frac{\partial a_n}{\partial\beta}
+
N_cN_fu_n\frac{\partial b_n}{\partial\beta}
\right). 
\label{QCD-S3-energy-lowT}
\end{eqnarray}
Note that the expressions \eqref{QCD-S3-saddle-point}, \eqref{QCD-S3-free-energy-2} and 
\eqref{QCD-S3-energy-lowT} are valid only for $T\le T_{\rm GWW}$.

\begin{figure}[htb]
\begin{center} 
\scalebox{1}{
\includegraphics[width=9cm]{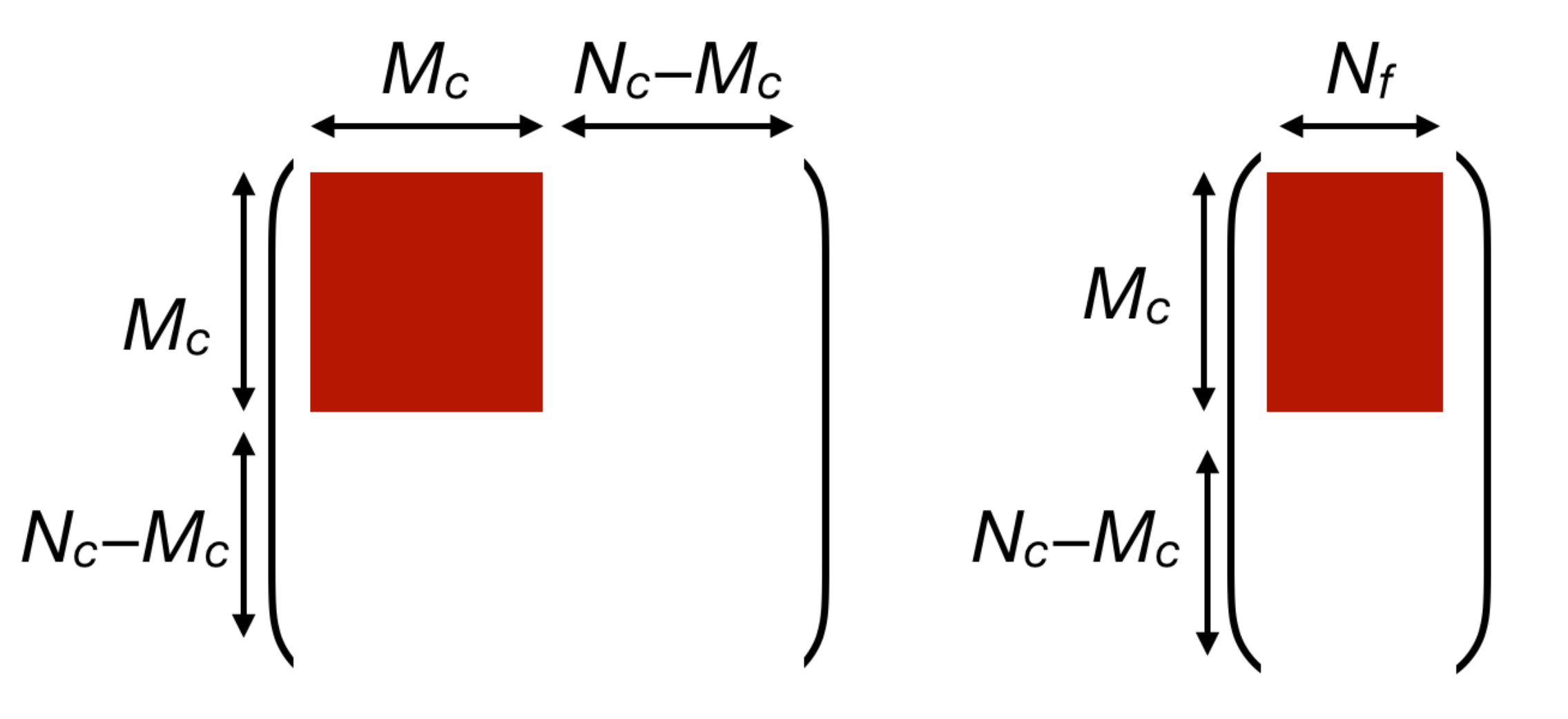}}
\scalebox{1}{
\includegraphics[width=9cm]{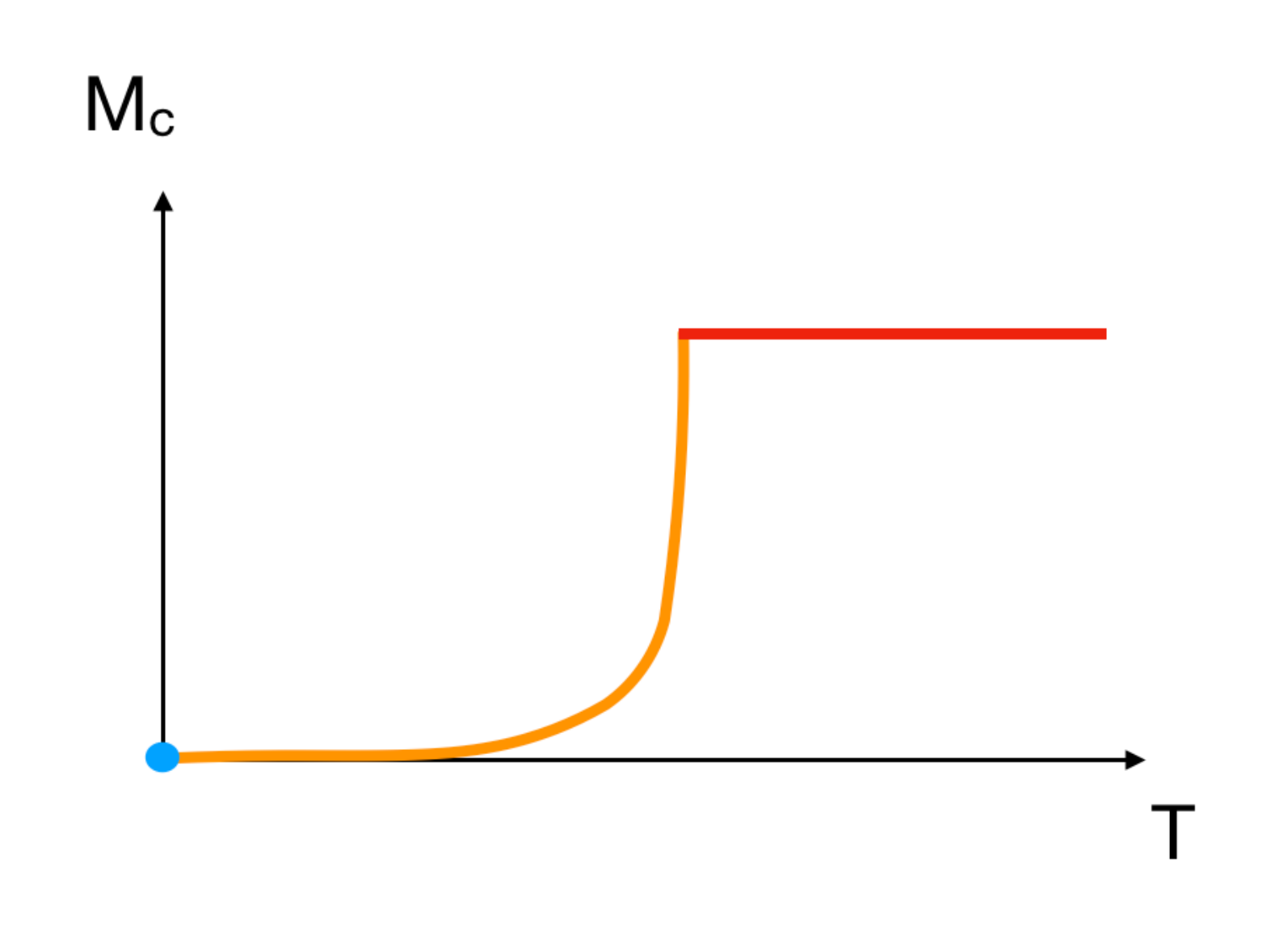}}
\end{center}
\caption{
Above is a cartoon of partial deconfinement in the weakly coupled QCD. 
In the gauge sector, SU($M_c$) block is deconfined, while the rest is confined. 
In the quark sector, $M_c$ colors of all $N_f$ flavors are deconfined. 
Complete confinement is realized only at $T=0$. 
}
\label{fig:partial_deconfinement_weakly_coupled_QCD}
\end{figure}

At this point, we need to demonstrate that partial deconfinement is actually taking place at a temperature $T\le T_{\rm GWW}$. 
To begin, consider the situation that $M_c$ colors and $N_f$ flavors are deconfined, 
as shown in Fig.~\ref{fig:partial_deconfinement_weakly_coupled_QCD}. 
Corresponding to a given temperature $T$, we identify the size of the deconfined block $M_c$ such that 
$T$ is the GWW-transition temperature of the SU($M_c$) theory, which we denote $T= T_{\rm GWW}(M_c,N_f)$. 

We start by computing free energy $\tilde{F}$ for an SU$(M_c)$ YM theory with $N_f$ fundamental fermions of mass $m$. This is trivially obtained by replacing $N_c$ in \eqref{QCD-S3-free-energy} by $M_c$, 
\begin{eqnarray}
\beta \tilde{F}
=
\sum_n\left(
M_c^2a_n(T)
\tilde{u}_n^2
+
M_cN_f
b_n(T)\tilde{u}_n
\right).
\label{QCD-S3-free-energy-Mc}
\end{eqnarray}
Here, for convenience in distinguishing variables, we have expressed the phase distribution of the Polyakov line in the SU$(M_c)$ theory as
\begin{eqnarray}
\tilde{\rho}(\theta) \equiv \frac{1}{2\pi}\left(1+ 2\sum_n \tilde{u}_n \cos(n\theta)\right). 
\end{eqnarray}
We can now determine the GWW temperature by solving 
\begin{eqnarray}
\min 
\tilde{\rho}(\theta) 
=
0. 
\label{distribution-SU(M)}
\end{eqnarray}
By identifying $\tilde{u}=\frac{N_c}{M_c}u$ we can trivially rewrite \eqref{QCD-S3-free-energy-Mc} to \eqref{QCD-S3-free-energy}, 
while \eqref{distribution-SU(M)} changes to 
\begin{eqnarray}
\min
\rho(\theta)
=
\frac{1}{2\pi}\left(1-\frac{M_c}{N_c}\right),
\label{how-to-determine-Mc}
\end{eqnarray}
where again $\rho(\theta)$, given in \eqref{SUNc-Polyakov-phase}, is the phase distribution of the Polyakov line in the SU($N_c$) theory. 
It is easy to see that the key relation \eqref{eq:Polyakov_loop_partial_deconfinement} is actually satisfied. 
This relation \eqref{how-to-determine-Mc} gives a simple way to identify $M_c$ for a given temperature. 
See Fig.~\ref{fig:partial_deconfinement_weakly_coupled_QCD} for a qualitative picture for the temperature dependence. 

Further, computing the energy $E$ in the SU$(M_c)$ theory straightforward and results in 
\begin{eqnarray}
E
&=&
\sum_n\left(
N_c^2u_n^2\frac{\partial a_n}{\partial\beta}
+
N_cN_fu_n\frac{\partial b_n}{\partial\beta}
\right)
\nonumber\\
&=&
\sum_n\left(
M_c^2\tilde{u}_n^2\frac{\partial a_n}{\partial\beta}
+
M_cN_f\tilde{u}_n\frac{\partial b_n}{\partial\beta},
\right),
\end{eqnarray}
and the entropy is $S=\beta(E-F)$ is easily obtained from the above expressions.  By substituting $T=T_{\rm GWW}(M_c,N_f)$, we obtain 
\begin{eqnarray}
E
=
E_{\rm GWW}(M_c,N_f),\qquad\text{and}\qquad S
=
S_{\rm GWW}(M_c,N_f). 
\end{eqnarray}
Therefore, the energy and entropy behave consistently with \eqref{how-to-determine-Mc}. 

In Ref.~\cite{Hanada:2019czd}, a method of constructing a one-to-one mapping between the states in the Hilbert spaces of the SU($N_c$) theory and the SU($M_c$) theory was discussed and it is completely applicable in the above analysis.
Likewise following  Ref.~\cite{Hanada:2019czd}, the gauge symmetry `spontaneously breaks' to SU($M_c$)$\times$SU($N_c-M_c$)$\times$U(1). 

In the original treatment of this theory \cite{Schnitzer:2004qt},  the GWW transition has been identified with the deconfinement transition.  The reasoning behind this identification is based on the observation that the free energy is of order $N_f^2$ below the GWW transition -- as can be seen in \eqref{QCD-S3-free-energy-2} -- and the argument that $N_f^2$ counts the number of mesonic degrees of freedom.  
However, as we have seen, the GWW transition should be identified with the deconfinement transition of all color degrees of freedom. Below this maximal GWW transition, there is a coexistence phase on the internal space, i.e. the group manifold of $SU(N_c)$, where confined and deconfined sectors are simultaneously realized for a complementary subset of the color degrees of freedom.

\subsection{Chiral symmetry}
\hspace{0.51cm}

In this section, our aim is to understand the fate of chiral symmetry as one moves into the partially deconfined phase.  That is, does an intermediate phase connecting phases of broken and unbroken chiral symmetry exist?  While there are no quantitative, dynamical calculations yet to determine the existence of such a coexistence phase, by exploiting a known loophole in the Vafa-Witten theorem \cite{Vafa:1983tf} we will explore the possibility that an intermediate phase with partially broken chiral symmetry could exist.

\begin{figure}[htb]
\begin{center} 
\scalebox{1}{
\includegraphics[width=9cm]{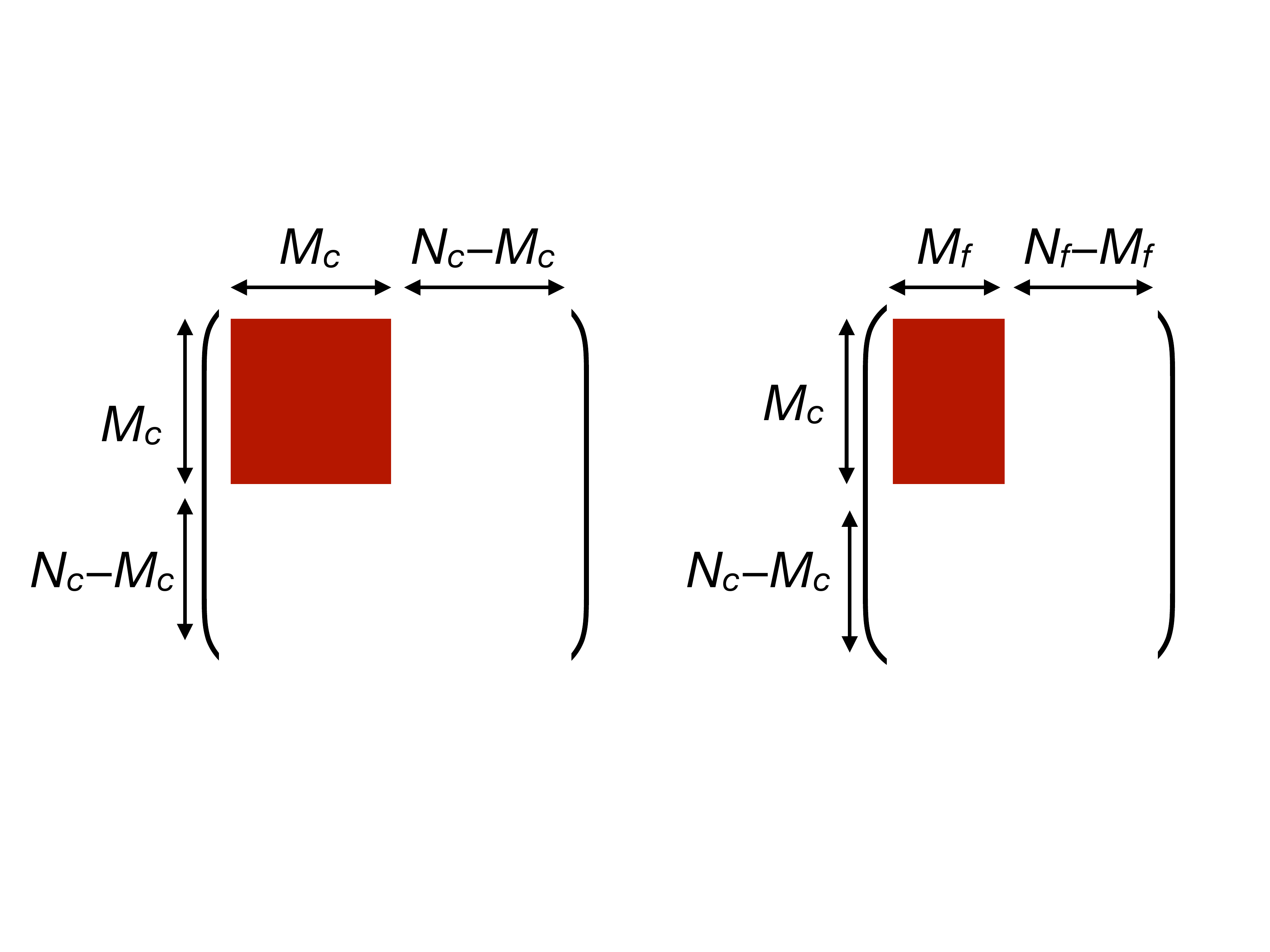}}
\end{center}
\caption{
In the gauge sector, SU($M_c$) block is deconfined, while the rest is confined. 
In the quark sector, $M_c$ colors of $M_f$ flavors are deconfined. 
The chiral symmetry is broken in the deconfined sector; namely, for $M_c$ colors it breaks to SU($M_f$)$_A$, 
while the other $N_c-M_c$ colors it breaks completely. 
Based on the Vafa-Witten theorem \cite{Vafa:1983tf}, $M_f=N_f$ is expected at zero baryon chemical potential.
}
\label{fig:gauge-chiral-synchronization}
\end{figure}

Let us now consider the infinite volume limit of large-$N_c$ SU($N_c)$ QCD with $N_f$ fermions in the fundamental representation.
In order to preserve the chiral symmetry precisely, let us consider quarks in this theory to be exactly massless.
We will write the massless quarks as $\psi_{cf}$ with the color and flavor indices taking values $c=1,2,\cdots,N_c$ and $f=1,2,\cdots,N_f$, respectively.
If an intermediate phase connecting the confined phase and deconfined phase does exist, a naive guess as to its structure could be as follows:
\begin{itemize}
\item
In the gauge sector, SU($M_c$) $\subset$ SU($N_c$) deconfines, and we assume that the symmetry breaking occurs in such a way that $c=1,2,\cdots,M_c$ spans the deconfined block as in the left panel of Fig.~\ref{fig:gauge-chiral-synchronization}.

\item
In the quark sector, $\psi_{cf}$ with $c=1,2,\cdots,M_c$ and $f=1,2,\cdots,M_f$ deconfines. That is, we have $M_c$ color and $M_f$ flavor degrees of freedom deconfined as depicted in the right panel of Fig.~\ref{fig:gauge-chiral-synchronization}. 
\item The left over ($N_c^2-M_c^2$ degrees of freedom in the gauge sector and $N_cN_f-M_cM_f$ degrees of freedom in the quark sector) remain confined.  
\end{itemize}

There are, however, immediate objections that could be raised if the above case is true. One should note above that the minimal flavor symmetry breaking pattern would involve the breaking of $SU(N_f)_V$.
Such flavor symmetry breaking patter would seem to be in violation of the Vafa-Witten theorem \cite{Vafa:1983tf},
which forbids the breaking of the vector part of the flavor symmetry.
However, we can use a well-know exception: 
the Vafa-Witten theorem assumes the positivity of the fermion determinant in the Euclidean path integral, which is dependent on the value of the baryon chemical potential $\mu_B$.  
At $\mu_B=0$, the above assumption is valid, and hence we expect $M_f=N_f$, 
just as Fig.~\ref{fig:partial_deconfinement_weakly_coupled_QCD}.
However, one can introduce $\mu_B\neq 0$ without breaking chiral symmetry, and because the fermion determinant is complex at finite baryon chemical potential, the Vafa-Witten theorem does not apply.  Thus, $M_f<N_f$ at $\mu\neq 0$ may be possible. 

If we can assume that the deconfinement pattern above holds, 
a natural expectation is that the chiral condensate is zero for $c=1,2,\cdots,M_c$ and $f=1,2,\cdots,M_f$.  
Then, in the SU($M_c$) sector, SU($N_f$)$_A$ spontaneously breaks to SU($M_f$)$_A$, 
and SU($N_f$)$_V$ spontaneously breaks to SU($M_f$)$_V\times$SU($N_f-M_f$)$_V$. 
In the SU($N_c-M_c$) sector, SU($N_f$)$_A$ breaks completely, and SU($N_f$)$_V$ does not break.  
Here, $M_c$ and $M_f$ are determined dynamically as the function of the energy. While we cannot analytically prove this statement, we can do a consistency check by using the 't Hooft anomaly, which is somewhat trivial.

In the usual sense of anomaly matching in the broken and unbroken chiral symmetry phases, treating the global chiral symmetry SU($N_f$)$_A$ as a background gauge symmetry the anomaly is easily calculated in the UV to be proportional to ${\rm dim}\ {\rm SU}(N_f) = N_f^2-1$.  If along the flow to the IR SU($N_f$)$_A$ is completely broken, the anomaly of the WZW action of the NG boson takes the same value. 

However, the logic of anomaly matching should be satisfied at any intermediate energy scale provided there are no accidental symmetries along the flow.  
Then, in the partially deconfined phase where there is an admixture deconfined and confined sectors coexisting we should be able to perform the same counting as above.
If the gauge symmetry and chiral symmetry break in a coordinated manner as we have conjectured above in the deconfined SU($M_c$) sector, then the quarks contribute $M_f^2-1$, while the NG boson corresponding to the coset SU($N_f$)$_A$/SU($M_f$)$_A$contribute $N_f^2-M_f^2$. Thus, the sum is always $N_f^2-1$ and the anomaly in the partially deconfined phase matches the UV anomaly.  
In the confined SU($N_c-M_c$) sector, the anomaly matching is trivially satisfied: The chiral symmetry is completely broken or unbroken.

As noted in the many caveats throughout this section, the breaking pattern we have discussed above is merely a logical possibility,
and we are without the dynamical calculations needed in order to see if it is actually realized. 
It would be interesting to understand the effect of the non-degenerate quark mass on the flavor symmetry breaking pattern.
Partial breaking of flavor symmetry may take place in other corners of the phase diagram as well, 
such as the hadron condensation at low temperature and large baryon chemical potential.  

\section{Pure Yang-Mills in two dimensions}\label{sec:2d-YM}
\hspace{0.51cm}
In this section, we consider two solvable examples of two-dimensional pure Yang-Mills theory in Euclidean spacetime at finite coupling: Wilson's plaquette action and the continuum theory on two-sphere.  
In these models, we will explicitly demonstrate that as the `t Hooft coupling $\lambda$ varies, a transition into a phase analogous to the coexistence phase of confined and deconfined degrees of freedom takes place. 
An important difference from the previous example is that these examples do not admit the Wick rotation to Minkowski spacetime, 
and thus, we are not manifestly concerned about the interpretation of states in the Hilbert space. 
Rather, we focus our attention on the properties of the saddle points dominating the path integral in the large-$N$ limit. 
We obtain simple results consistent with our picture of partial-symmetry-breaking phase transitions,
which supports our conjecture that such transitions are generic.    

\subsection{Gross-Witten-Wadia transition in 2d lattice gauge theory}\label{sec:2d-lattice}
\hspace{0.51cm}
The first model that we will consider is the two-dimensional version of Wilson's plaquette formulation of U($N$) gauge theory on a square lattice.  
This theory has been solved at large $N$, and a third order (GWW) transition has been found \cite{Gross:1980he,Wadia:2012fr}.  Historically, this model is, obviously, the first theory in which the GWW was found; the derivation of which goes as follows. 

We start by fixing the axial gauge. That is, we enforce that the link variable connecting lattice sites along the time direction is taken to be the unit matrix. 
In the axial gauge, the partition function becomes 
\begin{eqnarray}
Z
=
\left(
\int d\mathcal{U} e^{-\frac{1}{g^2}\left({\rm Tr}\mathcal{U}+{\rm Tr}\mathcal{U}^\dagger\right)}
\right)^V
\equiv z^V, \label{GWW-lattice-theory}
\end{eqnarray}
where $V$ is the number of lattice sites, the plaquette $\mathcal{U}$ is $N\times N$ unitary matrix, and the integral is taken with respect to the Haar measure. 
Therefore, the problem of computing the full path integral is reduced to simply computing the one-matrix model partition function $z$. 

To solve this theory, let $\{e^{i\theta_1},\,\ldots,\,e^{i\theta_N}\}$ denote the set of eigenvalues of $\mathcal{U}$ where the phases take values in $\theta_i\in (-\pi,\pi]$ for all $i=1,\,\ldots,\,N$. 
The large-$N$ limit is taken with fixed $\lambda=g^2N$. 
In the large $N$ limit, the one-matrix model can be solved by finding the saddle point 
expressed by the distribution of phases, $\rho(\theta)$, which has a non-trivial dependence on the coupling.

For $\lambda \ge 2$, the distribution of phases takes the form
\begin{eqnarray}
\rho(\theta)
=
\frac{1}{2\pi}\left(
1
+
\frac{2}{\lambda}\cos\theta
\right).\label{Lattice-strong-coupling-phase-distbn}
\end{eqnarray}
However, in the regime where $\lambda<2$ the distribution has a markedly different profile in $\theta$
\begin{eqnarray}
\rho(\theta)
=
\left\{
\begin{array}{cc}
\frac{2}{\pi\lambda}\cos\frac{\theta}{2}
\sqrt{\frac{\lambda}{2}-\sin^2\frac{\theta}{2}}
& (|\theta|\le 2\arcsin\sqrt{\frac{\lambda}{2}})\\
0 & (|\theta| > 2\arcsin\sqrt{\frac{\lambda}{2}})
\end{array}
\right..\label{Lattice-weak-coupling-phase-distbn}
\end{eqnarray}
The free energy $F=-\log z$ near $\lambda =2$ in the two distinct regimes described above displays a phase transition:
\begin{eqnarray}
\frac{F}{N^2}=
\left\{
\begin{array}{cc}
-\frac{1}{\lambda^2} & (\lambda\ge 2)\\
-\frac{2}{\lambda}+\frac{1}{2}\log\left(\frac{2}{\lambda}\right)+\frac{3}{4}& \lambda< 2
\end{array}
\right..
\end{eqnarray}
One can immediately deduce the existence of a third order phase transition at $\lambda =2$ insofar as at the transition point $F$, $\frac{dF}{d\lambda}$ and $\frac{d^2F}{d\lambda^2}$ are continuous, while $\frac{d^3F}{d\lambda^3}$ discontinuously jumps. This is the original version of the GWW transition. 

In the asymptotically free gauge theories which exhibit the confinement/deconfinement transition, because the coupling constant becomes smaller at higher temperature there is an immediate identification of the weakly coupled phase of the theory with deconfinement and the strongly coupled phase of the theory with the confinement.  
Hence, we are tempted to see the analogy between the GWW transition in the 2d lattice theory considered above and the confinement/deconfinement transition in higher dimensional gauge theories. 
Pushing this analogy, the distribution of the plaquette phases plays the role of the distribution of the Polyakov line phases.  
The analogue of the confined phase (i.e. $\rho(\theta)=\frac{1}{2\pi}$ for the Polyakov line phase distribution) is seen in comparing to \eqref{Lattice-strong-coupling-phase-distbn} at exactly infinite coupling $\lambda = \infty$. 
Then to complete the anaology for any finite coupling, $2<\lambda<\infty$, the plaquette phase distribution describes a phases of the lattice theory resembling the partially deconfined phase. 

Let us make this observation more precise. Returning to the sketch of partial deconfinement in an SU($N$) gauge theory from above, we assume a ``deconfined'', weakly coupled SU($M$) $\subset$ SU$(N)$ sector is sitting at the GWW point,
while the remaining degrees of freedom is in the ``confined'', strongly coupled limit. 
Pushing the analogy to its end, the distribution \eqref{eq:Polyakov_loop_partial_deconfinement}, 
compared to the expectation from the generic property of partial deconfinement \eqref{Lattice-strong-coupling-phase-distbn}, suggests the following identification: 
\begin{eqnarray}
\frac{M}{N}=\frac{2}{\lambda}  \qquad
\text{or equivalently}\qquad
\lambda=\frac{2N}{M}. \label{GWW-analogy-lambda}
\end{eqnarray}

Perhaps not surprisingly, the value of $\lambda$ in \eqref{GWW-analogy-lambda} actually corresponds to the GWW-point of the SU($M$) theory.  As discussed in the previous section, the 't Hooft coupling in the truncated  SU($M$) theory is $\lambda_{M}=g^2M=\frac{M}{N}\cdot\lambda$. We've just seen that the GWW transition in the lattice theory at $\lambda_M=2$, which maps to a transition at $\lambda=\frac{2N}{M}$ in this analogy. 

Making this analogy even more precise, it follows that in the at the GWW transition, modulo gauge transformations, a matrix of the following form dominates the path integral:
\begin{eqnarray}
\mathcal{U}
=
\left(
\begin{array}{cc}
\mathcal{U}_{M,{\rm GWW}} & \\
0 & \mathcal{U}_{N-M,\lambda=\infty}
\end{array}
\right).  
\label{GWW-model-symmetry-breaking}
\end{eqnarray}
Here, $\mathcal{U}_{M,{\rm GWW}}$ is the $M\times M$ unitary matrix contributing at the GWW point, 
and $\mathcal{U}_{N-M,\lambda=\infty}$ is the $(N-M)\times (N-M)$ unitary matrix contributing at the strong coupling limit.  
Like in the partially deconfined phase, it is convenient to think that the gauge symmetry is 
`spontaneously broken' to SU($M$)$\times$SU($N-M$)$\times$U(1).   Further with the above identification, one can see that at the GWW point the free energy computed from \eqref{GWW-lattice-theory} on the saddle in \eqref{GWW-model-symmetry-breaking} takes the same value as the free energy of the SU($M$)-truncated theory: 
\begin{eqnarray}
F(\lambda)=-\frac{N^2}{\lambda^2}=-\frac{M^2}{4}=F_{\rm GWW}(M),
\end{eqnarray}
where the form of the saddle point \eqref{GWW-model-symmetry-breaking} allows us to interpret it simply as the sum of the 
contributions from the SU($M$) and SU($N-M$) sectors, which are $F_{\rm GWW}(M)$ and zero, respectively.

\subsection{Douglas-Kazakov transition in 2d YM on S$^2$}
\hspace{0.51cm}
Next let us consider the large-$N$ two-dimensional U($N$) Yang-Mills theory on two-sphere S$^2$. 
Unlike Sec.~\ref{sec:2d-lattice}, we consider the theory at the continuum limit. 
The 't Hooft coupling $\lambda$ has the dimension of $({\rm mass})^2$, 
and nontrivial phase structure can be seen when the dimensionless combination $\lambda A$ is varied,
where $A$ is the area of S$^2$ and $\lambda=g_{\rm YM}^2N$ is the 't Hooft coupling.
There is a third order phase transition similar to the GWW transition, which is called the Douglas-Kazakov (DK) transition \cite{Douglas:1993iia}. 
We examine this transition following Ref.~\cite{Gross:1994mr}, and exhibit the partially-symmetry-breaking phase transition in the strong coupling region. 
This theory is slightly more nontrivial compared to the previous examples, in which a simple truncation to the SU($M$)-sector worked. 

The partition function is given by $Z(\lambda A,N)=\int [dA_\mu]e^{-S}$, 
where the action is the usual 2d Yang-Mills on S$^2$ with area $A$, 
\begin{eqnarray}
S
=
-\frac{N}{4\lambda}\int d^2x\sqrt{g(x)}{\rm Tr}F^{\mu\nu}F_{\mu\nu}. 
\end{eqnarray}
The free energy $F=-\log Z$ is given by 
\begin{eqnarray}
F(\lambda A,N)
=
\frac{\lambda A}{24}(N^2-1)
-
\sum_{n=1}^{N}
\log h_n\left(\frac{g_{\rm YM}^2 A}{2}\right),  
\label{eq:free-energy-2dYMonS2}
\end{eqnarray}
where\footnote{
$h_n$ here corresponds to $h_{n-1}$ in Ref.~\cite{Gross:1994mr}. 
}
\begin{eqnarray}
h_n\left(\frac{g_{\rm YM}^2 A}{2}\right)
=
\sqrt{2\pi}(n-1)!
\cdot \left(\frac{g_{\rm YM}^2 A}{2}\right)^{3/2-n}
\label{eq:h_n-weak-coupling}
\end{eqnarray}
if 
\begin{eqnarray}
A < \frac{\pi^2}{(n-1) g_{\rm YM}^2}, 
\end{eqnarray}
and more complicated form at $A > \frac{\pi^2}{(n-1) g_{\rm YM}^2}$. 

The form \eqref{eq:free-energy-2dYMonS2} is already very suggestive: 
$h_1,\cdots,h_M$ can be regarded as the contribution from the U($M$)-sector. 
As $\lambda A$ decreases from $\infty$, the DK transition takes place
when the weak-coupling form \eqref{eq:h_n-weak-coupling} becomes valid for all $n=1,2,\cdots,N$ \cite{Gross:1994mr}.  This is true in particular at the critical area $A_c(g_{\rm YM}^2,N)$ given by 
\begin{eqnarray}
A_c(g_{\rm YM}^2,N)=\frac{\pi^2}{g_{\rm YM}^2N}=\frac{\pi^2}{\lambda}.
\label{eq:DK-critical-area}
\end{eqnarray} 

As can be demonstrated easily from \eqref{eq:free-energy-2dYMonS2}, the phase transition being probed is third order and resembles the GWW transition in the previous examples. 
Thus, by interpreting $(\lambda A)^{-1}$ as `temperature', we expect that the strong- and weak-coupling phases are similar to 
the partially deconfined and completely deconfined phases.
Then, the natural expectation is that 
$h_1,h_2,\cdots,h_M$ ($M=\frac{\pi^2}{g_{\rm YM}^2A}$) are the contributions from the analog of the `deconfined' sector.
In this context then, $h_{M+1},\cdots,h_N$ are the analog of the `confined' sector contributions. 
In order for this interpretation to be consistent with the partial phase transition picture, 
the critical area of the U($M$) theory, $A_{\rm c}(g^2,M)$, has to satisfy the following consistency condition:
\begin{eqnarray}
M
=
\frac{\pi^2}{g_{\rm YM}^2 A_{\rm c}(g_{\rm YM}^2,M)}. 
\end{eqnarray}
It is a trivial exercise to check that this condition is actually satisfied in the above example.

There is, perhaps, an interesting extension of the above discussion to the quantum group version of 2d YM.  As has been noted in many places, e.g. \cite{Arsiwalla:2005jb, Jafferis:2005jd}, q-deformed YM on an S$^2$ exhibits a line of third-order (DK) phase transitions much like ordinary YM theory for $p\geq 2$.  Further, some of the intermediate phase behavior that is discussed above is hinted at explicitly in the extension of Fig. 3 in \cite{Jafferis:2005jd} to the q-deformed case, which raises an interesting question about the role of partial-symmetry-breaking phase transitions in, say, black holes and topological strings \cite{Aganagic:2004js}.  However, these are questions that we will leave for future analysis.

\section{Black hole/black string topology change and 2d SYM theory}\label{sec:BH-BS-2d-YM}
\hspace{0.51cm}
In this section, we consider 2d ${\cal N}=(8,8)$ SU($N$) SYM theory (i.e. with 16 supercharges).
We consider the Euclidean version of the theory, and put it  on a torus $T^2 = S^1_\beta \times S^1_L$ where $\beta$ and $L$ are the radii of the thermal and spatial circles, respectively. 
For the bosonic fields, the periodic boundary condition is imposed along both circles, 
while for the fermionic fields the periodic and anti-periodic boundary conditions are imposed along the spatial and temporal circles, respectively. 
With this boundary condition, $\beta$ is interpreted as the inverse of temperature. 
The action is obtained by dimensionally reducing 4d ${\cal N}=4$ SYM to 2d resulting in the action
\begin{eqnarray}
S
=
\frac{N}{\lambda_{\rm 2d}}
\int_0^\beta dt\int_0^L dx\ {\rm Tr}
\left(
\frac{1}{4}F_{\mu\nu}^2
+
\frac{1}{2}(D_\mu X_I)^2
-
\frac{1}{4}[X_I,X_J]^2
\right)
+
({\rm fermion\ part}), \label{eq:2d-max-SYM-bosonic-action}
\end{eqnarray}
where $X_I$ ($I=1,2,\cdots,8$) are scalar fields. 
There is a $\mathbb{Z}_{N_c}$ center symmetry, which acts on the Wilson line $W$ winding on spatial circle as 
$W\to e^{2\pi ik/N}W$, where $k$ is integer. 
We fix $L$ and vary energy $E$ or temperature $T= 1/\beta$. 
As we will see, as $E$ decreases, the SU($M$)-sector goes into in the center broken phase and $M$ gradually increases to $N$. 
\subsection{Relation to black hole/black string topology change}
\hspace{0.51cm}
Maximally supersymmetric 2d SU($N$) SYM theory captures features of the black hole/black string transition \cite{Gregory:1993vy,Kol:2002xz} as follows \cite{Aharony:2004ig}. 

Firstly, 2d $\mathcal{N}=(8,8)$ SYM theory arises in type IIB string theory as the worldvolume theory of $N$ D1-branes in $\mathbb{R}^{9,1}$. 
When the spatial dimension of the 2d theory is compactified, one of the nine spatial dimensions is compactified as well, and D1-branes wrap on the compactified dimension. 
T-dualizing along the worldvolume, the $N$ D1-branes are mapped to $N$ D0-branes, and the circumference of the compactified circle becomes $L'\equiv \frac{4\pi^2\alpha'}{L}$ . 

Staying in the T-dual picture and using the gauge/gravity correspondence, by varying $T\ll L^{-1}$ while keeping fixed the size of spatial circle $L$, the gravitational description of the D0-branes is a tiny black hole (black zero-brane). 
The size of the black hole is much smaller than $L'$, and hence the solution is well approximated by that in the noncompact space. 
As we increase the temperature, black hole becomes bigger, and eventually wraps on the spatial circle and turns to black string. 
The transition temperature is higher for higher values of  the (T-dual) spatial length scale $L'$ -- or equivalently the transition temperature goes up as $L$ is decreased.  

\begin{figure}[htbp]
\begin{center}
\scalebox{0.5}{\includegraphics{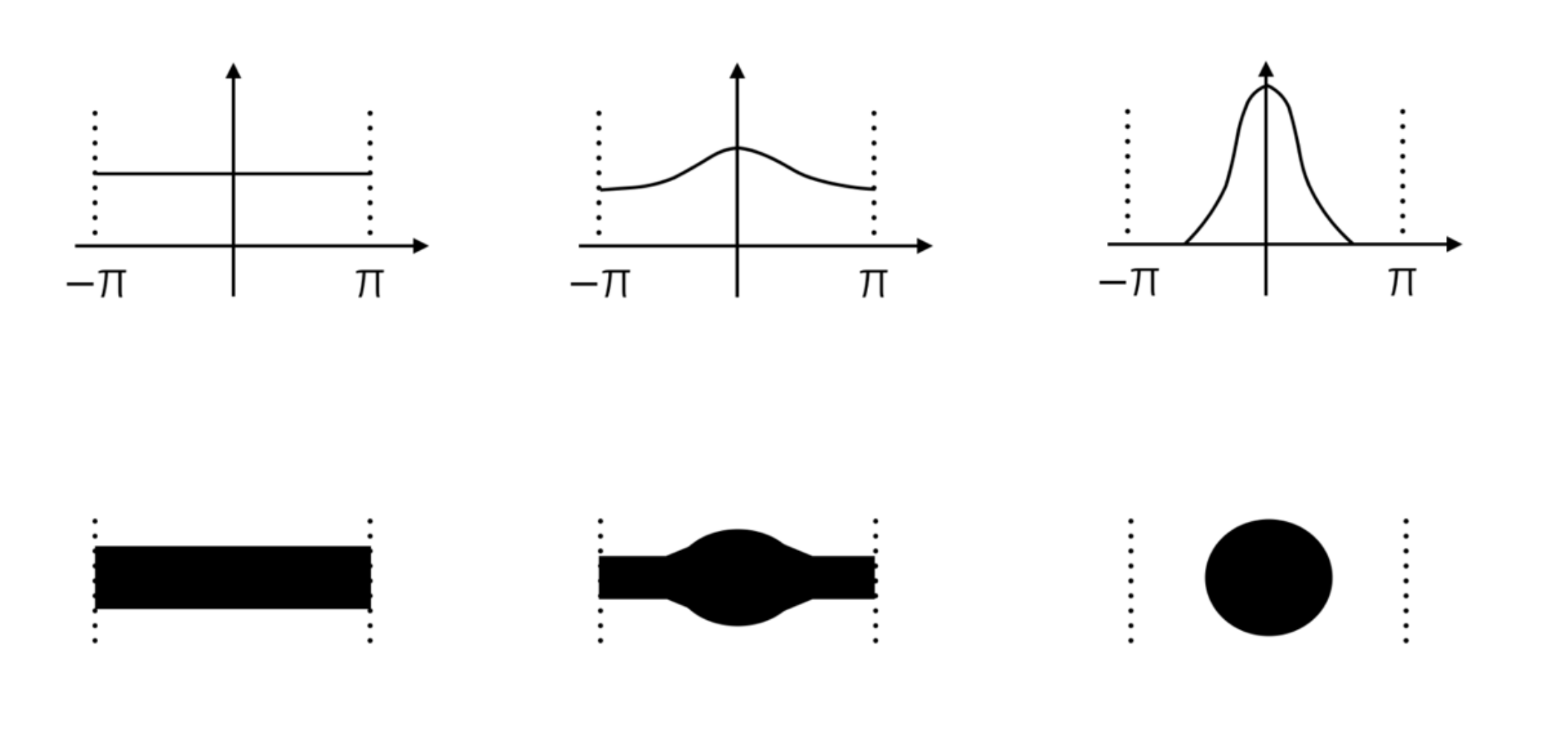}}
\end{center}
\caption{[Top] The distribution of Wilson line phases, uniform (left), non-uniform but not gapped (middle) and gapped (right). 
[Bottom] The counterparts in gravity side. Uniform black string (left), non-uniform black string (middle) and black hole (right). 
}\label{fig:2dSYM-phase-distribution}
\end{figure}

\begin{figure}[htbp]
\begin{center}
\scalebox{0.2}{\includegraphics{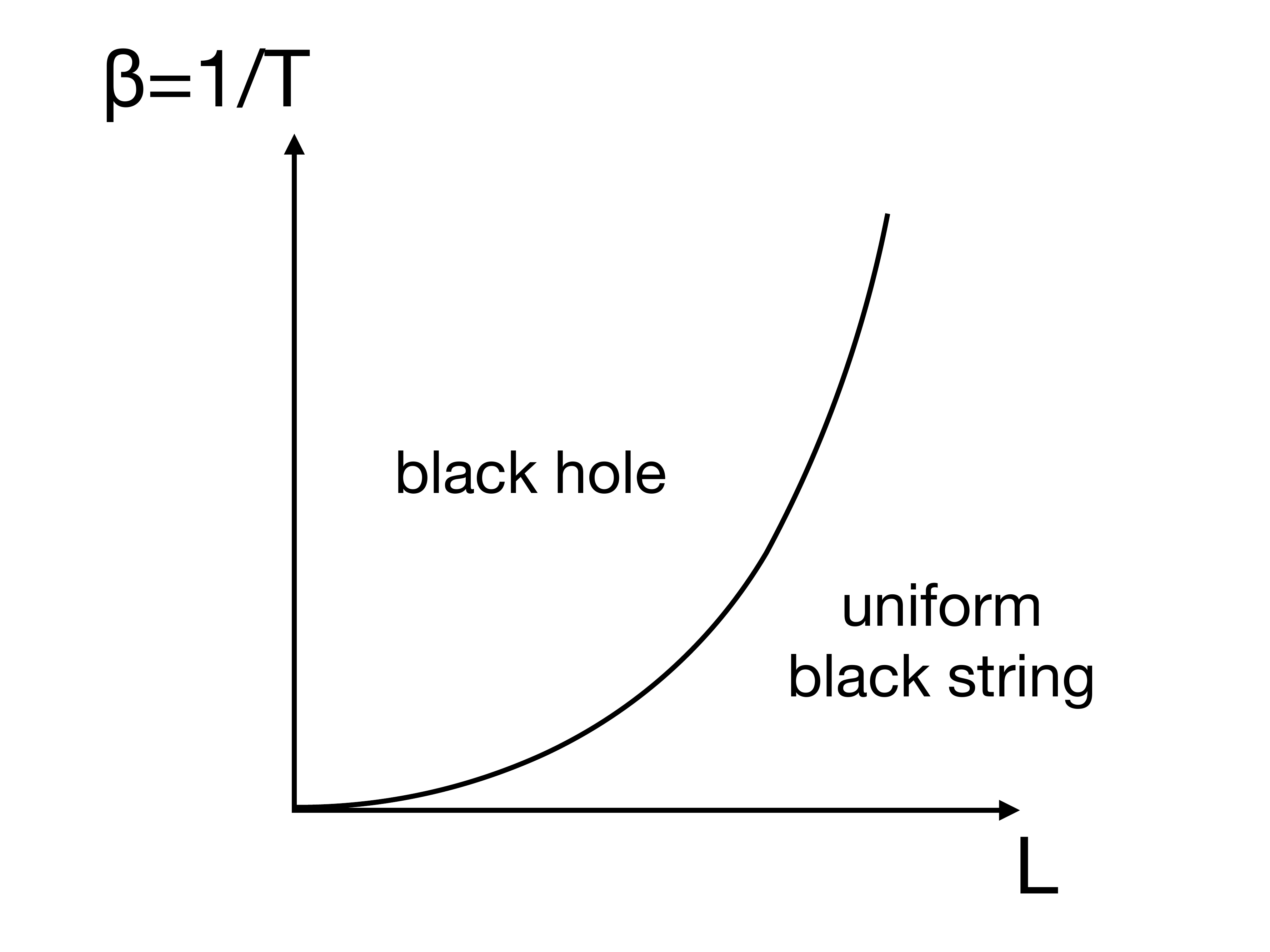}}
\end{center}
\caption{A cartoon picture of phase diagram of 2d maximal SYM on spatial circle. 
$L_{\rm c}\sim \beta^{1/3}$ at high temperature, 
$L_{\rm c}\sim \beta^{1/2}$ at low temperature.  
}\label{fig:2dSYM-phase-diagram}
\end{figure}

In terms of the SYM theory description, the locations of D0-branes are the phases of Wilson lines $W$ wrapped on the spatial S$^1$.
The distribution of Wilson line phases can be uniform, non-uniform but not gapped, or gapped; see Fig.~\ref{fig:2dSYM-phase-distribution}. 
Their gravitational descriptions are the uniform black string, non-uniform black string and black hole, respectively. The phase diagram is shown in Fig.~\ref{fig:2dSYM-phase-diagram}. In the following sections, we will detail the phase transition and find the intermediate phase. 
 
\subsection{Small volume, high temperature region}
\hspace{0.51cm}
While the simulation of 2d SU($N$) SYM theory is costly for arbitrary $N$ at generic temperatures,\footnote{
See Refs.~\cite{Catterall:2010fx,Catterall:2017lub} for serious attempts and nice summary of gauge/gravity duality in this system. } at high temperature and small volume (i.e. $\beta,\,L \rightarrow 0$)
the analysis simplifies as follows \cite{Aharony:2004ig,Aharony:2005ew}. 
Let us start with Euclidean path integral of the theory on $T^2= S^1_\beta\times S^1_L$ described above.  
Imposing thermal boundary condition the fermions are anti-periodic along the temporal circle, while bosons are periodic. Along the spatial circle, we impose periodic boundary conditions on both bosons and fermions.

 At high temperature, $\beta=\frac{1}{T} \rightarrow 0$, the fermions are gapped out due to the large Kaluza-Klein mass and decouple.  Integrating out the fermionic degrees of freedom and taking care of zero modes leaves behind the dimensionally reduced 1d bosonic matrix model, which is a good description of the physics near $\beta = 0$. 
 
Explicitly, after the dimensional reduction the action is given by 
\begin{eqnarray}
S_{\rm 1d}
=
\frac{N}{\lambda_{\rm 2d}T}\int_0^L dx {\rm Tr}\left(
\frac{1}{2}(D_xX_I)^2
-
\frac{1}{4}
[X_I,X_J]^2
\right).  \label{SYM-reduced-MQM}
\end{eqnarray} 
The bosonic degrees of freedom in the dimensionally reduced theory, $X_I (I=1,2,\cdots,9)$, are $N\times N$ Hermitian matrices, and the gauge covariant derivative is $D_xX_I=\partial_xX_I-i[A_x,X_I]$, where $A_x$ is the remaining component the gauge field.   By rescaling $\tilde{x}=(\lambda_{\rm 2d}T)^{1/3}x$, $\tilde{L}=(\lambda_{\rm 2d}T)^{1/3}L$,  $\tilde{D}_x=(\lambda_{\rm 2d}T)^{-1/3}D_x$ and  $\tilde{X}_I=(\lambda_{\rm 2d}T)^{-1/3}X_I$, we can rewrite \eqref{SYM-reduced-MQM} as 
\begin{eqnarray}
S_{\rm 1d}
=
N\int_0^{\tilde{L}} d\tilde{x} {\rm Tr}\left(
\frac{1}{2}(\tilde{D}_x\tilde{X}_I)^2
-
\frac{1}{4}
[\tilde{X}_I,\tilde{X}_J]^2
\right).  \label{SYM-reduced-MQM-rescaled}
\end{eqnarray} 
In \eqref{SYM-reduced-MQM-rescaled}, when one calculates the expectation values of the form $\langle f(\tilde{X})\rangle=\frac{1}{Z}\int [d\tilde{X}]f(\tilde{X})e^{-S}$, 
the only parameter is $\tilde{L}=(\lambda_{\rm 2d}T)^{1/3}L$, and hence,
large volume and high temperature are equivalent. 

This high-temperature/large volume equivalence is seen in the sketched phase diagram in Fig.~\ref{fig:2dSYM-phase-diagram}: If one starts in the black hole phase, one can see the transition to the black string phase by either by going to larger $T$ with fixed $L$ or larger $L$ with fixed $T$. The critical value of $L$ at fixed temperature scales as $L_{\rm c}\sim (\lambda_{\rm 2d}T)^{-1/3}$. 

\begin{figure}[htbp]
\begin{center}
\includegraphics[width=0.5\textwidth]{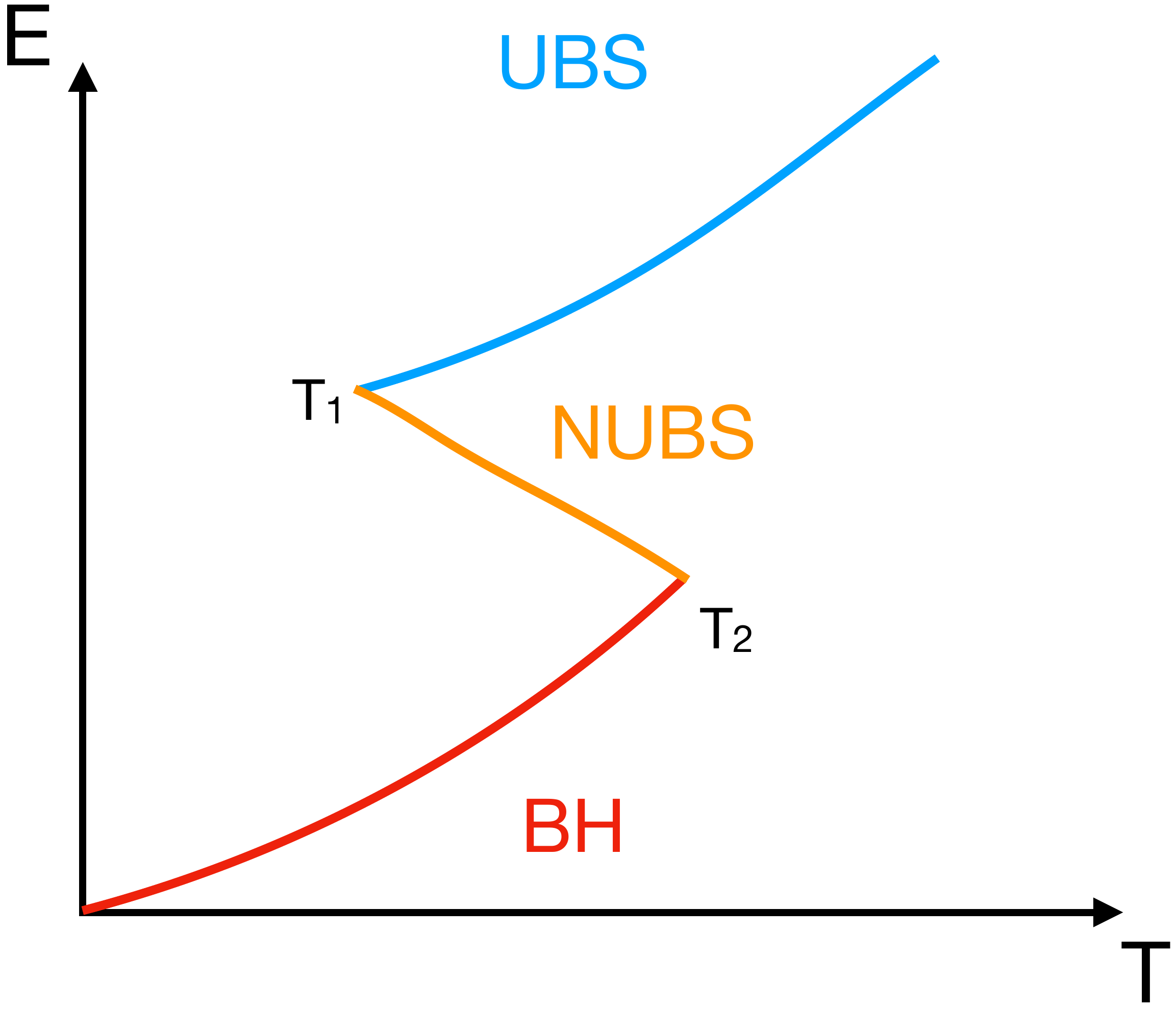}
\end{center}
\caption{A cartoon picture of 
the black hole/black string transition in the microcanonical ensemble, at high-T (weak coupling). 
Red: black hole (BH), blue: uniform black string (UBS), orange: non-uniform black string (NUBS). 
}\label{fig:BH-BS-microcanonical-high-T}
\end{figure}

The computational advantage afforded by the high temperature limit is that \eqref{SYM-reduced-MQM-rescaled} can more easily be studied numerically using standard Markov chain Monte Carlo (MCMC) methods (see e.g.~ Refs.\cite{Hanada:2018fnp,DeGrand:2019boy} for introductory articles). 
In the MCMC paradigm, the authors of Ref.~\cite{Bergner:2019rca} treated $\tilde{L}$ as `inverse temperature' of the dimensionally reduced theory and observed a first order transition 
with hysteresis caused by the unstable saddle of the free energy (see the left panels of Fig.~\ref{fig:E-vs-T-BH} and Fig.~\ref{fig:M-vs-T-BH}).  
In the 2d theory, we should see the transition of the same type when we change $T$ instead of $\tilde{L}$. 
Hence the intermediate phase --- non-uniform black string --- should have negative specific heat. 
Therefore, we expect the microcanonical phase diagram for fixed small $L$ shown in Fig.~\ref{fig:BH-BS-microcanonical-high-T}.  
In the microcanonical description, the non-uniform black string phase is stable, because there is only one maximum of entropy at each value of $E$. 
This is consistent with the classical real-time simulation \cite{Hanada:2018qpf} which observed the stable non-uniform black string phase.
Note that a phase with negative specific heat corresponds to a maximum of the free energy in the canonical ensemble. 

\subsection{Large volume, low temperature region}
\begin{figure}[htbp]
\begin{center}
\includegraphics[width=0.5\textwidth]{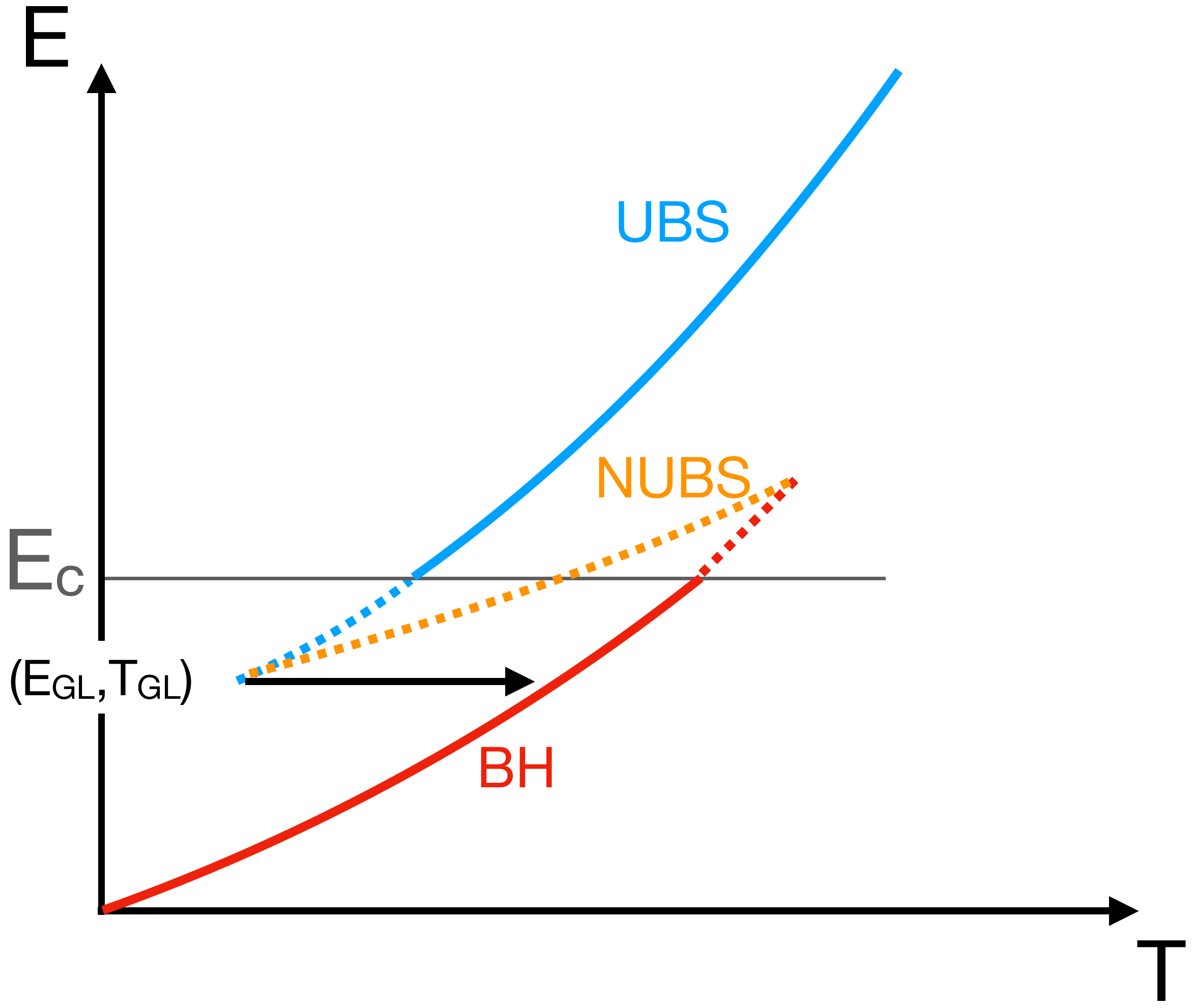}
\end{center}
\caption{
Black hole/black string transition in the microcanonical ensemble. 
Red: black hole (BH), blue: uniform black string (UBS), orange: non-uniform black string (NUBS). 
Solid line: entropy global maximum, dotted line: local maximum (but not global maximum). 
The black arrow show the Gregory-Laflamme instability. 
}\label{fig:BH-BS-microcanonical}
\end{figure}

\hspace{0.51cm}
When $L$ is sufficiently large ($L\gg\frac{1}{\sqrt{\lambda_{\rm 2d}}}$), the transition takes place when the supergravity description of the T-dual (D0-brane) picture is a good approximation \cite{Aharony:2004ig}. 
A simplified picture of the plausible microcanonical ensemble is shown in 
Fig.~\ref{fig:BH-BS-microcanonical} with the caveat that the non-uniform string phase can be more complicated;\footnote{
There has been a long debate regarding this picture, especially regarding the final state of the Gregory-Laflamme instability \cite{Gregory:1993vy} (see also Sec.~\ref{sec:GL-instability-disappear?}). 
Ref.~\cite{Horowitz:2001cz} discussed that the black string cannot pinch off, at least in classical gravity, then the final state is likely to be the non-uniform black string.    
However Ref.~\cite{Wiseman:2002zc} studied a non-uniform string solution, and fond that the entropy is smaller than that of uniform black string, which suggested that the non-uniform string is not the final state.  
Numerical simulation \cite{Choptuik:2003qd} observed the emergence of array of black holes connected by thin black strings, but the simulation failed before the topology change to take place due to the large gradient. 
Refs.~\cite{Kudoh:2004hs,Kleihaus:2006ee,Headrick:2009pv,Kalisch:2017bin} observed that black hole and non-uniform black string meet at the topology change point, as shown in Fig.~\ref{fig:BH-BS-microcanonical}. 
By now various solutions with the S$^1$ compactification have been investigated and rich phase structure has been realized, see e.g.~\cite{Dias:2007hg,Figueras:2012xj,Emparan:2014pra,Dias:2017uyv}. 
} there may be multiple solutions. 
Compared to the phase diagram in the high temperature limit in Fig.~\ref{fig:BH-BS-microcanonical-high-T},
the structure of saddle points in intermediate energy scale is richer in that multiple phases can coexist.  
Above the critical value $E_{\rm c}$, uniform black string is entropically favored, and below $E_{\rm c}$ the black hole geometry is favored. 
At the intermediate energy scale, there exist other less entropically favorable saddles that are none-the-less stable in the large-$N$ limit. 

If we begin in the stable uniform-black string phase and gradually lower the energy, 
the black string remains stable until the energy reaches the Gregory-Laflamme point $E_{\rm GL}$.  At $E_{\rm GL}$, the black string entropy ceases to be locally maximum and dynamical instability to long wavelength perturbations sets in \cite{Gregory:1993vy}.  In the same manner, starting in the stable black hole phase and gradually increasing the energy, the black hole remains stable until its entropy ceases to be local maximum and becomes  unstable to small perturbations.  When either of the initial phases reach a point at which they become entropically disfavored and instability sets in, another stable branch appears --- non-uniform black string --- and bridges the black hole and uniform black string phases. 
In this sense, the microcanonical phase diagram is still continuous despite the existence of the first order transition at $E=E_{\rm c}$. 

Different from the high temperature limit, now the non-uniform black string phase has the positive specific heat. Thus, 
the non-uniform black string corresponds to a local minimum of the free energy in the canonical ensemble. 
In certain temperature range, there are three stable phases, even in the canonical ensemble\footnote{Strictly speaking, two of the phases are metastable. However, in the large-$N$ limit the tunneling is parametrically suppressed as the probability of such an event is $\mathcal{O}(e^{-N^2})$.}. 

\subsection{Non-uniform black string as partial-symmetry-breaking phase}
\hspace{0.51cm}
The intermediate, non-uniform black string phase can be regarded as the partial-symmetry-breaking phase in which the SU($N$) gauge symmetry is `spontaneously broken' to 
SU($M$)$\times$SU($N-M$). An important point to note in this phase is that the $\mathbb{Z}_N$ center symmetry is completely broken, and the realized $\mathbb{Z}_{N-M}$ center symmetry is \textit{not} a subgroup of the $\mathbb{Z}_N$ symmetry.

One natural question is as follows. 
We claimed that the specific heat is negative in the non-uniform black string phase at high-temperature and small-volume, 
and it exists stably in microcanonical ensemble. 
Why is it possible? Why don't we have the coexistence of two phases separated in volume? 
To understand it, firstly recall the argument in Sec.~\ref{sec:large-volume-vs-small-volume}: at small volume, 
whether the thermodynamic instability sets in depends on the detail of the dynamics. 
The case under consideration is the center symmetry breaking phase transition, which is essentially the onset of the decoupling of the KK mode. 
After the KK modes decouple, the system has to be uniform along the spatial circle, and hence, cannot occupy a small fraction of the physical space. 

\subsection{Stringy effects eliminate the Gregory-Laflamme instability?}\label{sec:GL-instability-disappear?}
\hspace{0.51cm}
It is interesting to see how stringy corrections can affect the Gregory-Laflamme instability of the uniform black string \cite{Gregory:1993vy}. 
The Gregory-Laflamme instability means the instability of the uniform black string against a small perturbation, 
which is observed in classical gravity. It is believed that the black string eventually pinches off and become black hole; 
although a more complete understanding of quantum gravity is needed to resolve the singularity associated with the topology change. 

In terms of dual gauge theory, it is the instability expressed by the black arrow in Fig.~\ref{fig:BH-BS-microcanonical}. 
Note that this is a first order transition in the microcanonical ensemble. 
At high-temperature and small-volume, however, there seems to be no first order transition in the microcanonical ensemble
(Fig.~\ref{fig:BH-BS-microcanonical-high-T}). 
Maybe a small jump exists near the border between the uniform black string and non-uniform black string, or the one between 
the non-uniform black string and black hole, but at least the transition from uniform black string to black hole is impossible. 
It suggests that a large stringy correction eventually eliminates the Gregory-Laflamme instability. 

Note that a similar transition in the canonical ensemble survives. Namely, if the temperature is gradually lowered, 
a jump from black string to black hole at the same temperature but smaller energy takes place.

\section{Yang-Mills with adjoint fermions on ${\mathbb R}^3\times$S$^1$}\label{sec:R3*S1_pbc}
\hspace{0.51cm}
Let us consider another example where it is very easy to see the gauge symmetry breaking: 
SU($N_c$) Yang-Mills theory with $N_f$ Majorana adjoint fermions of mass $m$ on ${\mathbb R}^3\times$S$^1$.  For the example below, we will denote the radius of the S$^1$ by $L$ and consider cases where $N_f\ge 2$.  
We impose periodic boundary conditions both for the gauge field and adjoint fermions around the S$^1$. 
When $L$ is small, the dimensionally reduced 3d picture is good, and the gauge field along the S$^1$ can be regarded as an adjoint Higgs field. 
At $L\ll m^{-1}$, the usual spontaneous symmetry breaking takes place, as we will see below. 

Staying in the regime of fixed small $L$ and large $m$, the phase diagram of this theory can be analyzed (to some extent) by using the 1-loop effective action \cite{Unsal:2010qh}
\begin{eqnarray}
V[\Omega]
=
\frac{2}{\pi^2L^4}\sum_{n=1}^\infty\left\{
-1
+
\frac{n_f}{2}(nLm)^2K_2(nLm)
\right\}
\frac{|{\rm Tr}~\Omega^n|^2}{n^4}
\equiv
\sum_{n=1}^\infty f_n(Lm)|{\rm Tr}~\Omega^n|^2, \label{eq:adjoint-qcd-effective-action}
\end{eqnarray}
where $\Omega ={\rm diag}(e^{i\theta_1},\cdots,e^{i\theta_N})$ is the Wilson line wrapping on the spatial circle and $K_2(z)$ is the modified Bessel function of the second kind. The free energy is obtained by minimizeing the effective action with respect to $\Omega$

Let $z=z_0$ be the solution of $-1+\frac{n_f}{2}(z)^2K_2(z)=0$. 
Since $z^2 K_2(z)$ is monotonically decreasing in the region $z>0$, at $\frac{z_0}{m}<L$, all $f_j(Lm)$'s are negative. Thus, the minimum of the effective action
is obtained by setting $|{\rm Tr}~\Omega^n|=1$ for any $n$, or equivalently $\theta_1=\theta_2=\cdots\theta_N$.
This solution to the saddle point equation is the $\mathbb{Z}_1$-center-symmetry phase (i.e. the center symmetry is completely broken), 
and correspondingly the solution preserves the full SU($N$) gauge symmetry. 

If we start to dial the size of the S$^1$ such that $L$ becomes smaller, the coefficients $f_n(Lm)$ turns positive at $L<\frac{z_0}{nm}$, and the center symmetry is restored as \cite{Unsal:2010qh}
\begin{eqnarray}
\mathbb{Z}_1\to \mathbb{Z}_2\to \mathbb{Z}_3\to\cdots.
\nonumber 
\end{eqnarray}
\noindent
The center symmetry restoration coincides with the gauge symmetry breaking pattern
\begin{eqnarray}
{\rm SU}(N)
\to
\left[{\rm SU}(N/2)\right]^2\times {\rm U}(1)
\to 
\left[{\rm SU}(N/3)\right]^3\times\left[{\rm U}(1)\right]^2
\to\cdots. 
\nonumber 
\end{eqnarray} 
Apparently, an intermediate phase must exist between $\mathbb{Z}_k$- and $\mathbb{Z}_{k+1}$-center-symmetric phases. 

In order to understand the appearance and meaning of the intermediate phase, let us start by considering the simplest case $\mathbb{Z}_1\to \mathbb{Z}_2$.  In terms of gauge symmetry breaking, we want to study the phase connecting SU($N$) and SU$(N/2)\times$SU$(N/2)$, which naively will interpolate SU$(N-M)\times$SU$(M)$ from $M=0$ to $M=N/2$.  

We start with a simple ansatz of a block diagonal form for the holonomy matrix in the intermediate phase: 
\begin{eqnarray}
\Omega&=&\begin{pmatrix} \mathds{1}_{\rm N-M}& 0 \\0 &-\mathds{1}_{\rm M}\end{pmatrix},
\end{eqnarray}
where $\mathds{1}_d$ is a the $d$-dimensional unit matrix. It is straightforward, then to show that on this ansatz ${\rm Tr}~\Omega^n=N$ for even $n$ and ${\rm Tr}~\Omega^n=N-2M$ for odd $n$. Evaluating the effective action on the ansatz gives
\begin{eqnarray}
V[\Omega]
&=&
\sum_{n:{\rm even}}f_n(Lm)N^2
+
\sum_{n:{\rm odd}}f_n(Lm)
(N-2M)^2
\nonumber\\
&=&
\left(1-\frac{2M}{N}\right)^2V_1
+
\left(
1-\left(1-\frac{2M}{N}\right)^2
\right)V_2,  
\end{eqnarray}
where $V_1$ and $V_2$ are the free energies of the $\mathbb{Z}_1$ and $\mathbb{Z}_2$ configurations, respectively.  
When $V_1=V_2$ (namely $\sum_{n:{\rm odd}}f_n(Lm)=0$), the free energy does not depend on $M$, and hence, 
the first order transition without hysteresis appears.

To understand other intermediate phases that appear in the sequence $\mathbb{Z}_k\to\mathbb{Z}_{k+1}$, let us consider $\mathbb{Z}_2\to \mathbb{Z}_3$, which is still relatively simple to discuss and contains the salient features to understand generic $k$.  
As an intermediate phase, there is an SU$(M)$$\times$SU$(M)$$\times$SU$(N-2M)$$\times$U$(1)^2$-phase.  The holonomy matrix should then take the form
\begin{align}
\Omega=\begin{pmatrix}
 \phi ~\mathds{1}_{M} & 0 &0\\
 0& \phi^*~\mathds{1}_M & 0\\
 0 & 0 &\mathds{1}_{N-2M}
\end{pmatrix}.
\end{align}  
Note that $\phi=i$ when $M=N/2$ and $\phi=e^{2\pi i/3}$ when $M=N/3$. 
At each $M$, $\phi$ can be evaluated by extremizing the effective action within this ansatz.  
It is straightforward, though tedious, to generalize the argument above to generic $\mathbb{Z}_k\to \mathbb{Z}_{k+1}$ intermediate phases. 
Note that Ref.~\cite{Myers:2009df} studied several finite-$N$ theories and found analogous phases which are stable. 

\section{Discussions}
\hspace{0.51cm}
In this paper, we have examined a variety of theories with large rank symmetry groups and observed partial-symmetry-breaking phase transitions appear to be generic large $N$ phenomena leading to interesting phase coexistence behavior.
In each case, the partial-symmetry-breaking phase appears as an intermediate phase connecting two (meta-)stable phases that generically realize the full rank symmetry group. 
While the examples we have discussed are gauge theories, the underlying mechanism discussed in Sec.~\ref{sec:qualitative_argument}
does not fundamentally require a gauge symmetry at all.  This leads somewhat naturally to our speculation that such partial symmetry breaking phase transitions could admit a broad universality class. After all, the demonstration of these novel transitions through matrix models certainly suggests that theories that have limiting descriptions in terms of random matrices could exhibit partial symmetry in intermediate phases as well. However, it is unclear how to construct a generic proof of universality, and so it would be worthwhile to explore possible examples that lie outside of the scope of gauge theories.  
 
While we have focussed exclusively on examples in gauge theories here (apart from the toy model of ant trails), there are a number of ways in which partial symmetry breaking phase transitions could appear in other areas of physics.  Indeed, there seem to be natural analogs for these transitions observed in the theories we have studied above in phenomena in statistical physics models such as quantum rotors, Dimer models, and Bose-Einstein condensates that could be well-described by the same language.  While it is not immediately clear what, if any, immediate utility this reinterpretation could have, it may be worth trying to understand the connections between confinement/deconfinment and phase transitions in condensed matter that could afford a more unified picture of the fundamental physics governing them.

Returning to gauge theoretic examples, one example where one might expect to possibly see partial symmetry breaking phase transitions would be in a particular deformation of the index (or rather the supersymmetric partition function on S$^1\times\rm{S}^{d-1}$) of superconformal field theories.
It was noted in \cite{Choi:2018vbz} that in analytically continuing fugacities, $x$, in the superconformal index of 4d $\mathcal{N}=4$ SU($N$) SYM theory, one could observe a confinement/deconfinement transition by taking $x$ to be a complex number.  
It was previously argued that unlike the partition function, supersymmetric indices are invariant under changes in continuous parameters, and so phase transitions would not be captured by any index computation \cite{Kinney:2005ej}.  It would be interesting to see if this deformed index does see any partial symmetry breaking phase transitions between the `confined' and `deconfined' phases; if such partial symmetry breaking phase transitions are visible in the index with complex fugacities and given the vast literature on supersymmetric and superconformal indices, it would also be interesting to see any broader implications. 

It is also interesting to note that we have seen in Sec.~\ref{sec:free-QCD}, partial deconfinement can happen in QCD-like theories. 
A consequence of the partial-symmetry-breaking phase transition is that a simultaneous breaking and restoration of gauge symmetry is involved, 
which may give a precise definition of confinement/deconfinement transition based on the change of symmetry \cite{Hanada:2019czd}. 
(Note that it is widely believed that there is no symmetry characterizing deconfinement, because center symmetry is explicitly broken 
by the existence of quarks in the fundamental representation.)
It is important to understand whether the phenomenon we have discovered survives at $N_c=3$ and infinite volume, 
and whether there are consequences in observable quantities.

\begin{center}
\section*{Acknowledgements}
\end{center}
\hspace{0.51cm}
We thank Tom Cohen, Oscar Dias, Nick Evans, Pau Figueras, Antal Jevicki, Tatsuhiro Misumi, Andy O'Bannon, Cheng Peng, 
Masaki Tezuka, Hiromasa Watanabe, Nico Wintergerst, Toby Wiseman and Naoki Yamamoto for stimulating discussions and comments. 
The work of MH was partially supported by the STFC Ernest Rutherford Grant ST/R003599/1 and JSPS  KAKENHI  Grants17K1428. The work of BR was partially supported through the STFC Consolidated Grant ST/L000296/1 and by the KU Leuven C1 grant ZKD1118 C16/16/005.

\bibliographystyle{utphys}
\bibliography{partial-deconfinement-examples}

\providecommand{\href}[2]{#2}\begingroup\raggedright\begin{thebibliography}{10}

\bibitem{Hanada:2016pwv}
M.~Hanada and J.~Maltz, ``{A proposal of the gauge theory description of the
  small Schwarzschild black hole in AdS$_5\times$S$^5$},''
  \href{http://dx.doi.org/10.1007/JHEP02(2017)012}{{\em JHEP} {\bfseries 02}
  (2017) 012},
\href{http://arxiv.org/abs/1608.03276}{{\ttfamily arXiv:1608.03276 [hep-th]}}.

\bibitem{Berenstein:2018lrm}
D.~Berenstein, ``{Submatrix deconfinement and small black holes in AdS},''
  \href{http://dx.doi.org/10.1007/JHEP09(2018)054}{{\em JHEP} {\bfseries 09}
  (2018) 054},
\href{http://arxiv.org/abs/1806.05729}{{\ttfamily arXiv:1806.05729 [hep-th]}}.

\bibitem{Hanada:2018zxn}
M.~Hanada, G.~Ishiki, and H.~Watanabe, ``{Partial Deconfinement},''
  \href{http://dx.doi.org/10.1007/JHEP03(2019)145}{{\em JHEP} {\bfseries 03}
  (2019) 145},
\href{http://arxiv.org/abs/1812.05494}{{\ttfamily arXiv:1812.05494 [hep-th]}}.

\bibitem{Hanada:2019czd}
M.~Hanada, A.~Jevicki, C.~Peng, and N.~Wintergerst, ``{Anatomy of
  Deconfinement},''
\href{http://arxiv.org/abs/1909.09118}{{\ttfamily arXiv:1909.09118 [hep-th]}}.

\bibitem{Dai:1989ua}
J.~Dai, R.~G. Leigh, and J.~Polchinski, ``{New Connections Between String
  Theories},''
\href{http://dx.doi.org/10.1142/S0217732389002331}{{\em Mod. Phys. Lett.}
  {\bfseries A4} (1989) 2073--2083}.

\bibitem{Witten:1995im}
E.~Witten, ``{Bound states of strings and p-branes},''
  \href{http://dx.doi.org/10.1016/0550-3213(95)00610-9}{{\em Nucl. Phys.}
  {\bfseries B460} (1996) 335--350},
\href{http://arxiv.org/abs/hep-th/9510135}{{\ttfamily arXiv:hep-th/9510135
  [hep-th]}}.

\bibitem{Sundborg:1999ue}
B.~Sundborg, ``{The Hagedorn transition, deconfinement and N=4 SYM theory},''
  \href{http://dx.doi.org/10.1016/S0550-3213(00)00044-4}{{\em Nucl. Phys.}
  {\bfseries B573} (2000) 349--363},
\href{http://arxiv.org/abs/hep-th/9908001}{{\ttfamily arXiv:hep-th/9908001
  [hep-th]}}.

\bibitem{Aharony:2003sx}
O.~Aharony, J.~Marsano, S.~Minwalla, K.~Papadodimas, and M.~Van~Raamsdonk,
  ``{The Hagedorn - deconfinement phase transition in weakly coupled large N
  gauge theories},'' \href{http://dx.doi.org/10.4310/ATMP.2004.v8.n4.a1}{{\em
  Adv. Theor. Math. Phys.} {\bfseries 8} (2004) 603--696},
  \href{http://arxiv.org/abs/hep-th/0310285}{{\ttfamily arXiv:hep-th/0310285
  [hep-th]}}.
[,161(2003)].

\bibitem{Berkowitz:2016znt}
E.~Berkowitz, M.~Hanada, and J.~Maltz, ``{Chaos in Matrix Models and Black Hole
  Evaporation},'' \href{http://dx.doi.org/10.1103/PhysRevD.94.126009}{{\em
  Phys. Rev.} {\bfseries D94} no.~12, (2016) 126009},
\href{http://arxiv.org/abs/1602.01473}{{\ttfamily arXiv:1602.01473 [hep-th]}}.

\bibitem{Beekman9703}
M.~Beekman, D.~J.~T. Sumpter, and F.~L.~W. Ratnieks, ``Phase transition between
  disordered and ordered foraging in pharaoh{\textquoteright}s ants,''
  \href{http://dx.doi.org/10.1073/pnas.161285298}{{\em Proceedings of the
  National Academy of Sciences} {\bfseries 98} no.~17, (2001) 9703--9706},
  \href{http://arxiv.org/abs/http://www.pnas.org/content/98/17/9703.full.pdf}{{\ttfamily
  http://www.pnas.org/content/98/17/9703.full.pdf}}.
  \url{http://www.pnas.org/content/98/17/9703}.

\bibitem{Schnitzer:2004qt}
H.~J. Schnitzer, ``{Confinement/deconfinement transition of large N gauge
  theories with N(f) fundamentals: N(f)/N finite},''
  \href{http://dx.doi.org/10.1016/j.nuclphysb.2004.06.057}{{\em Nucl. Phys.}
  {\bfseries B695} (2004) 267--282},
\href{http://arxiv.org/abs/hep-th/0402219}{{\ttfamily arXiv:hep-th/0402219
  [hep-th]}}.

\bibitem{Hollowood:2012nr}
T.~J. Hollowood and J.~C. Myers, ``{Deconfinement transitions of large N QCD
  with chemical potential at weak and strong coupling},''
  \href{http://dx.doi.org/10.1007/JHEP10(2012)067}{{\em JHEP} {\bfseries 10}
  (2012) 067},
\href{http://arxiv.org/abs/1207.4605}{{\ttfamily arXiv:1207.4605 [hep-th]}}.

\bibitem{Gross:1980he}
D.~J. Gross and E.~Witten, ``{Possible Third Order Phase Transition in the
  Large N Lattice Gauge Theory},''
\href{http://dx.doi.org/10.1103/PhysRevD.21.446}{{\em Phys. Rev.} {\bfseries
  D21} (1980) 446--453}.

\bibitem{Wadia:2012fr}
S.~R. Wadia, ``{A Study of U(N) Lattice Gauge Theory in 2-dimensions},''
\href{http://arxiv.org/abs/1212.2906}{{\ttfamily arXiv:1212.2906 [hep-th]}}.

\bibitem{Vafa:1983tf}
C.~Vafa and E.~Witten, ``{Restrictions on Symmetry Breaking in Vector-Like
  Gauge Theories},''
\href{http://dx.doi.org/10.1016/0550-3213(84)90230-X}{{\em Nucl. Phys.}
  {\bfseries B234} (1984) 173--188}.

\bibitem{Douglas:1993iia}
M.~R. Douglas and V.~A. Kazakov, ``{Large N phase transition in continuum QCD
  in two-dimensions},''
  \href{http://dx.doi.org/10.1016/0370-2693(93)90806-S}{{\em Phys. Lett.}
  {\bfseries B319} (1993) 219--230},
\href{http://arxiv.org/abs/hep-th/9305047}{{\ttfamily arXiv:hep-th/9305047
  [hep-th]}}.

\bibitem{Gross:1994mr}
D.~J. Gross and A.~Matytsin, ``{Instanton induced large N phase transitions in
  two-dimensional and four-dimensional QCD},''
  \href{http://dx.doi.org/10.1016/S0550-3213(94)80041-3}{{\em Nucl. Phys.}
  {\bfseries B429} (1994) 50--74},
\href{http://arxiv.org/abs/hep-th/9404004}{{\ttfamily arXiv:hep-th/9404004
  [hep-th]}}.

\bibitem{Arsiwalla:2005jb}
X.~Arsiwalla, R.~Boels, M.~Marino, and A.~Sinkovics, ``{Phase transitions in
  q-deformed 2-D Yang-Mills theory and topological strings},''
  \href{http://dx.doi.org/10.1103/PhysRevD.73.026005}{{\em Phys. Rev.}
  {\bfseries D73} (2006) 026005},
\href{http://arxiv.org/abs/hep-th/0509002}{{\ttfamily arXiv:hep-th/0509002
  [hep-th]}}.

\bibitem{Jafferis:2005jd}
D.~Jafferis and J.~Marsano, ``{A DK phase transition in q-deformed Yang-Mills
  on S**2 and topological strings},''
\href{http://arxiv.org/abs/hep-th/0509004}{{\ttfamily arXiv:hep-th/0509004
  [hep-th]}}.

\bibitem{Aganagic:2004js}
M.~Aganagic, H.~Ooguri, N.~Saulina, and C.~Vafa, ``{Black holes, q-deformed 2d
  Yang-Mills, and non-perturbative topological strings},''
  \href{http://dx.doi.org/10.1016/j.nuclphysb.2005.02.035}{{\em Nucl. Phys.}
  {\bfseries B715} (2005) 304--348},
\href{http://arxiv.org/abs/hep-th/0411280}{{\ttfamily arXiv:hep-th/0411280
  [hep-th]}}.

\bibitem{Gregory:1993vy}
R.~Gregory and R.~Laflamme, ``{Black strings and p-branes are unstable},''
  \href{http://dx.doi.org/10.1103/PhysRevLett.70.2837}{{\em Phys. Rev. Lett.}
  {\bfseries 70} (1993) 2837--2840},
\href{http://arxiv.org/abs/hep-th/9301052}{{\ttfamily arXiv:hep-th/9301052
  [hep-th]}}.

\bibitem{Kol:2002xz}
B.~Kol, ``{Topology change in general relativity, and the black hole black
  string transition},''
  \href{http://dx.doi.org/10.1088/1126-6708/2005/10/049}{{\em JHEP} {\bfseries
  10} (2005) 049},
\href{http://arxiv.org/abs/hep-th/0206220}{{\ttfamily arXiv:hep-th/0206220
  [hep-th]}}.

\bibitem{Aharony:2004ig}
O.~Aharony, J.~Marsano, S.~Minwalla, and T.~Wiseman, ``{Black hole-black string
  phase transitions in thermal 1+1 dimensional supersymmetric Yang-Mills theory
  on a circle},'' \href{http://dx.doi.org/10.1088/0264-9381/21/22/010}{{\em
  Class. Quant. Grav.} {\bfseries 21} (2004) 5169--5192},
\href{http://arxiv.org/abs/hep-th/0406210}{{\ttfamily arXiv:hep-th/0406210
  [hep-th]}}.

\bibitem{Catterall:2010fx}
S.~Catterall, A.~Joseph, and T.~Wiseman, ``{Thermal phases of D1-branes on a
  circle from lattice super Yang-Mills},''
  \href{http://dx.doi.org/10.1007/JHEP12(2010)022}{{\em JHEP} {\bfseries 12}
  (2010) 022},
\href{http://arxiv.org/abs/1008.4964}{{\ttfamily arXiv:1008.4964 [hep-th]}}.

\bibitem{Catterall:2017lub}
S.~Catterall, R.~G. Jha, D.~Schaich, and T.~Wiseman, ``{Testing holography
  using lattice super-Yang-Mills theory on a 2-torus},''
  \href{http://dx.doi.org/10.1103/PhysRevD.97.086020}{{\em Phys. Rev.}
  {\bfseries D97} no.~8, (2018) 086020},
\href{http://arxiv.org/abs/1709.07025}{{\ttfamily arXiv:1709.07025 [hep-th]}}.

\bibitem{Aharony:2005ew}
O.~Aharony, J.~Marsano, S.~Minwalla, K.~Papadodimas, M.~Van~Raamsdonk, and
  T.~Wiseman, ``{The Phase structure of low dimensional large N gauge theories
  on Tori},'' \href{http://dx.doi.org/10.1088/1126-6708/2006/01/140}{{\em JHEP}
  {\bfseries 01} (2006) 140},
\href{http://arxiv.org/abs/hep-th/0508077}{{\ttfamily arXiv:hep-th/0508077
  [hep-th]}}.

\bibitem{Hanada:2018fnp}
M.~Hanada, ``{Markov Chain Monte Carlo for Dummies},''
\href{http://arxiv.org/abs/1808.08490}{{\ttfamily arXiv:1808.08490 [hep-th]}}.

\bibitem{DeGrand:2019boy}
T.~DeGrand, ``{Lattice methods for students at a formal TASI},'' in {\em
  {Theoretical Advanced Study Institute in Elementary Particle Physics: The
  Many Dimensions of Quantum Field Theory (TASI 2019) Boulder, CO, USA, June
  3-28, 2019}}.
\newblock 2019.
\newblock
\href{http://arxiv.org/abs/1907.02988}{{\ttfamily arXiv:1907.02988 [hep-th]}}.
\newblock

\bibitem{Bergner:2019rca}
G.~Bergner, N.~Bodendorfer, M.~Hanada, E.~Rinaldi, A.~Schafer, and P.~Vranas,
  ``{Thermal phase transition in Yang-Mills matrix model},''
\href{http://arxiv.org/abs/1909.04592}{{\ttfamily arXiv:1909.04592 [hep-th]}}.

\bibitem{Hanada:2018qpf}
M.~Hanada and P.~Romatschke, ``{Real Time Quantum Gravity Dynamics from
  Classical Statistical Yang-Mills Simulations},''
  \href{http://dx.doi.org/10.1007/JHEP01(2019)201}{{\em JHEP} {\bfseries 01}
  (2019) 201},
\href{http://arxiv.org/abs/1808.08959}{{\ttfamily arXiv:1808.08959 [hep-th]}}.

\bibitem{Horowitz:2001cz}
G.~T. Horowitz and K.~Maeda, ``{Fate of the black string instability},''
  \href{http://dx.doi.org/10.1103/PhysRevLett.87.131301}{{\em Phys. Rev. Lett.}
  {\bfseries 87} (2001) 131301},
\href{http://arxiv.org/abs/hep-th/0105111}{{\ttfamily arXiv:hep-th/0105111
  [hep-th]}}.

\bibitem{Wiseman:2002zc}
T.~Wiseman, ``{Static axisymmetric vacuum solutions and nonuniform black
  strings},'' \href{http://dx.doi.org/10.1088/0264-9381/20/6/308}{{\em Class.
  Quant. Grav.} {\bfseries 20} (2003) 1137--1176},
\href{http://arxiv.org/abs/hep-th/0209051}{{\ttfamily arXiv:hep-th/0209051
  [hep-th]}}.

\bibitem{Choptuik:2003qd}
M.~W. Choptuik, L.~Lehner, I.~Olabarrieta, R.~Petryk, F.~Pretorius, and
  H.~Villegas, ``{Towards the final fate of an unstable black string},''
  \href{http://dx.doi.org/10.1103/PhysRevD.68.044001}{{\em Phys. Rev.}
  {\bfseries D68} (2003) 044001},
\href{http://arxiv.org/abs/gr-qc/0304085}{{\ttfamily arXiv:gr-qc/0304085
  [gr-qc]}}.

\bibitem{Kudoh:2004hs}
H.~Kudoh and T.~Wiseman, ``{Connecting black holes and black strings},''
  \href{http://dx.doi.org/10.1103/PhysRevLett.94.161102}{{\em Phys. Rev. Lett.}
  {\bfseries 94} (2005) 161102},
\href{http://arxiv.org/abs/hep-th/0409111}{{\ttfamily arXiv:hep-th/0409111
  [hep-th]}}.

\bibitem{Kleihaus:2006ee}
B.~Kleihaus, J.~Kunz, and E.~Radu, ``{New nonuniform black string solutions},''
  \href{http://dx.doi.org/10.1088/1126-6708/2006/06/016}{{\em JHEP} {\bfseries
  06} (2006) 016},
\href{http://arxiv.org/abs/hep-th/0603119}{{\ttfamily arXiv:hep-th/0603119
  [hep-th]}}.

\bibitem{Headrick:2009pv}
M.~Headrick, S.~Kitchen, and T.~Wiseman, ``{A New approach to static numerical
  relativity, and its application to Kaluza-Klein black holes},''
  \href{http://dx.doi.org/10.1088/0264-9381/27/3/035002}{{\em Class. Quant.
  Grav.} {\bfseries 27} (2010) 035002},
\href{http://arxiv.org/abs/0905.1822}{{\ttfamily arXiv:0905.1822 [gr-qc]}}.

\bibitem{Kalisch:2017bin}
M.~Kalisch, S.~Mockel, and M.~Ammon, ``{Critical behavior of the black
  hole/black string transition},''
  \href{http://dx.doi.org/10.1007/JHEP08(2017)049}{{\em JHEP} {\bfseries 08}
  (2017) 049},
\href{http://arxiv.org/abs/1706.02323}{{\ttfamily arXiv:1706.02323 [gr-qc]}}.

\bibitem{Dias:2007hg}
O.~J.~C. Dias, T.~Harmark, R.~C. Myers, and N.~A. Obers, ``{Multi-black hole
  configurations on the cylinder},''
  \href{http://dx.doi.org/10.1103/PhysRevD.76.104025}{{\em Phys. Rev.}
  {\bfseries D76} (2007) 104025},
\href{http://arxiv.org/abs/0706.3645}{{\ttfamily arXiv:0706.3645 [hep-th]}}.

\bibitem{Figueras:2012xj}
P.~Figueras, K.~Murata, and H.~S. Reall, ``{Stable non-uniform black strings
  below the critical dimension},''
  \href{http://dx.doi.org/10.1007/JHEP11(2012)071}{{\em JHEP} {\bfseries 11}
  (2012) 071},
\href{http://arxiv.org/abs/1209.1981}{{\ttfamily arXiv:1209.1981 [gr-qc]}}.

\bibitem{Emparan:2014pra}
R.~Emparan, P.~Figueras, and M.~Martinez, ``{Bumpy black holes},''
  \href{http://dx.doi.org/10.1007/JHEP12(2014)072}{{\em JHEP} {\bfseries 12}
  (2014) 072},
\href{http://arxiv.org/abs/1410.4764}{{\ttfamily arXiv:1410.4764 [hep-th]}}.

\bibitem{Dias:2017uyv}
O.~J.~C. Dias, J.~E. Santos, and B.~Way, ``{Localised and nonuniform thermal
  states of super-Yang-Mills on a circle},''
  \href{http://dx.doi.org/10.1007/JHEP06(2017)029}{{\em JHEP} {\bfseries 06}
  (2017) 029},
\href{http://arxiv.org/abs/1702.07718}{{\ttfamily arXiv:1702.07718 [hep-th]}}.

\bibitem{Unsal:2010qh}
M.~Unsal and L.~G. Yaffe, ``{Large-N volume independence in conformal and
  confining gauge theories},''
  \href{http://dx.doi.org/10.1007/JHEP08(2010)030}{{\em JHEP} {\bfseries 08}
  (2010) 030},
\href{http://arxiv.org/abs/1006.2101}{{\ttfamily arXiv:1006.2101 [hep-th]}}.

\bibitem{Myers:2009df}
J.~C. Myers and M.~C. Ogilvie, ``{Phase diagrams of SU(N) gauge theories with
  fermions in various representations},''
  \href{http://dx.doi.org/10.1088/1126-6708/2009/07/095}{{\em JHEP} {\bfseries
  07} (2009) 095},
\href{http://arxiv.org/abs/0903.4638}{{\ttfamily arXiv:0903.4638 [hep-th]}}.

\bibitem{Choi:2018vbz}
S.~Choi, J.~Kim, S.~Kim, and J.~Nahmgoong, ``{Comments on deconfinement in
  AdS/CFT},''
\href{http://arxiv.org/abs/1811.08646}{{\ttfamily arXiv:1811.08646 [hep-th]}}.

\bibitem{Kinney:2005ej}
J.~Kinney, J.~M. Maldacena, S.~Minwalla, and S.~Raju, ``{An Index for 4
  dimensional super conformal theories},''
  \href{http://dx.doi.org/10.1007/s00220-007-0258-7}{{\em Commun. Math. Phys.}
  {\bfseries 275} (2007) 209--254},
\href{http://arxiv.org/abs/hep-th/0510251}{{\ttfamily arXiv:hep-th/0510251
  [hep-th]}}.

\end{thebibliography}\endgroup

\end{document}